\documentclass{aa}  

\usepackage{graphicx}
\usepackage[varg]{txfonts}
\usepackage{lipsum}
\usepackage{subcaption}         
\usepackage{lscape}             
\usepackage{placeins}           
                                
\usepackage{physics}
                                
\usepackage[colorlinks=true,
		linkcolor=blue,
		citecolor=blue,
		urlcolor=blue]{hyperref}

\usepackage{natbib}
\bibpunct{(}{)}{;}{a}{}{,} 

\usepackage{multirow}

\usepackage{listings}

\usepackage{ulem}

\usepackage[dvipsnames]{xcolor} 

\definecolor{codegreen}{rgb}{0,0.6,0}
\definecolor{codegray}{rgb}{0.5,0.5,0.5}
\definecolor{codepurple}{rgb}{0.58,0,0.82}
\definecolor{backcolour}{rgb}{0.95,0.95,0.92}
\definecolor{darkgreen}{rgb}{0,0.8,0}

\newcommand{\model}[1]{\mbox{\textsf{#1}}}

\begin{document}

   \title{The Influence of the Fractal Dimension on Dust Evolution in Protoplanetary Disks}

   \author{J. E. Schöll\inst{\ref{ita}}
   		\and C. P. Dullemond\inst{\ref{ita}}
        \and C. Dominik\inst{\ref{api}}
        }

   \institute{Institut für Theoretische Astrophysik, Zentrum für Astronomie der Universität Heidelberg, Albert-Ueberle-Str. 2, 69120 Heidelberg, Germany \label{ita}
        \and Anton Pannekoek Institute for Astronomy, University of Amsterdam, Science Park 904, 1098 XH Amsterdam, The Netherlands \label{api}}

   \date{Received xxxx}

   \abstract
   {During the first stages of dust coagulation in protoplanetary disks, the dust aggregates are expected to have a high degree of porosity. Most models of dust growth, however, do not take this into account. The reason for this is the technical complexity of this problem. Furthermore, the coagulation/fragmentation kernel for colliding porous or fractal dust aggregates is not well understood.}
   {We wish to explore the effect of aggregate porosity on the evolution of the dust population in protoplanetary disks, with an emphasis on the fragmentation and the bouncing barrier.} 
   {We use the DustPy code, and implement porosity as a prescribed function of particle mass with the fractal dimension as a free parameter. In this way, we parameterize the ill-constrained physics of colliding porous/fractal aggregates, and we can explore the effect of different porosity prescriptions. We take into account the effect of porosity on the dust dynamics, while neglecting its effect on the collision outcomes.}
   {We find that larger particle masses are reached for lower fractal dimensions. The maximum Stokes numbers that are reached do not depend on the fractal dimension in the case of fragmentation-limited growth and decrease with decreasing fractal dimension in the case of bouncing-limited growth. Furthermore, particle growth is slower for smaller fractal dimensions in our models.}
   {The dust evolution is strongly influenced by the fractal dimension. Although larger masses are reached for smaller fractal dimensions, the particles are still much smaller than planetesimals. Under the assumption that the bouncing/fragmentation velocity does not depend on the fractal dimension or filling factor, fractal growth is not beneficial for the streaming instability to occur in the case of fragmentation-limited growth and even disadvantageous in the case of bouncing-limited growth.}


   \maketitle
   \nolinenumbers

\section{Introduction}
The coagulation of dust in protoplanetary disks is the first step along the way of forming planets and planetary systems. It has been studied extensively for decades, but progress is slow because realistic simulations require enormous resources, laboratory measurements are scarce, and observational constraints are only very indirect. As a result, the process of dust growth is still not sufficiently well understood.

What we do know is that collisional growth of dust aggregates encounters several so-called "growth barriers", such as the drift barrier \citep[e.g.,][]{Weidenschilling1977,Brauer2008}; bouncing barrier \citep[e.g.,][]{BlumWurm2008,Zsom2010,DominikDullemond2024}; fragmentation barrier \citep[e.g.,][]{Birnstiel2012}; and the electrostatic barrier \citep[e.g.,][]{Okuzumi2009a}. To produce larger bodies, mechanisms such as the streaming instability \citep{YoudinGoodman2005,JohansenYoudin2007,Johansen2007} are necessary. For the streaming instability to set in, the dust aggregates must grow to some minimal size, or more precisely, minimal Stokes number. Planetesimal formation is thus a two-stage process: first growth of dust aggregates to "pebble size", which are then collected into planetesimals by the streaming instability. Where and when grain growth manages to reach the right conditions for this to occur depends on the details of the grain growth process, which are hard to model.

In this paper we study how dust aggregate porosity, expressed in terms of a fractal growth model, affects the evolution of the dust distribution in a protoplanetary disk. To explain the challenges of this problem, and our method to address them, we first briefly review the algorithms used in modern dust evolution models.

The computational modeling of grain growth falls in one of two classes. One class of models simulates individual collisions of two dust aggregates, each consisting of a multitude of monomers, and follows its outcome \citep[e.g.,][]{DominikTielens1997,Suyama2008,Wada2007,Wada2008,Wada2009,Wada2011}. These models determine if the collision leads to sticking, bouncing, or fragmentation, and how the structure of the aggregates changes in the collision. These are the most detailed kind of dust coagulation models, and they can be directly compared to laboratory experiments \citep[e.g.,][]{Schraepler2012}. The other class of models simulates the evolution of a population of dust particles. While numerous approximate methods have been proposed \citep[e.g.,][]{EstradaCuzzi2008, Birnstiel2012}, two popular classes of models attempt to compute the evolution of the full grain size distribution: The Monte Carlo approach \citep[e.g.,][]{Ormel2007, ZsomDullemond2008, BeutelDullemond2023}, and the Smoluchowski equation approach \citep[][]{Weidenschilling1997, Tanaka2005, DullemondDominik2005}.

In the Monte Carlo method, the very large number of true particles (\(\sim 10^{30}\) or more, for \(\mathrm{\mu m}\)-sized particles) is represented by a computationally feasible number of computational particles. The advantage of this approach is that each computational particle can easily be assigned an arbitrary number of auxiliary properties (porosity, charge, composition etc.). The Monte Carlo method has been used to study the effect of porosity on the size distribution \citep[e.g.,][]{Ormel2007,Zsom2010,Krijt2015,Lorek2018,Xiang2020}, which demonstrated its importance.

In the approach that uses the Smoluchowski equation \citep{Smoluchowski1916}, the particle size distribution is computed directly in the form \(N(m,t)\), where \(N(m,t)\mathrm{d}m\) is the number of particles with masses between \(m\) and \(m+\mathrm{d}m\) at time \(t\). To do so, the mass coordinate is divided into \({\sim} 100\) logarithmically-spaced bins between the smallest and the largest dust aggregate mass. The advantage of this method is the higher dynamic range in mass, and the possibility to use implicit integration schemes to evolve the distribution over long time scales. For global disk models, the Smoluchowski approach is typically more efficient than the Monte Carlo approach. However, the disadvantage is that any additional independent property, such as the volume filling factor \(\phi\), adds another dimension, which often is computationally too expensive. A solution to this problem would be to introduce the volume filling factor \(\phi\) as dependent variable \(\phi(m,t)\). Models of this type \citep{Okuzumi2012,Estrada2022,Michoulier2024} remain efficient, while still being able to treat porosity in some detail. This requires a treatment of how porosity changes with time in the coagulation process, and as a result of, for instance, gas drag or self-gravity.

In the present paper, we choose to take a slightly simpler approach, in which we prescribe $\phi(m)$ beforehand, so that we do not need to compute it on-the-fly. The advantage of this simplification is that we can explore the effects of, for instance, a fractal dimension, without having to settle on a detailed model of how porosity is changing as a function of the evolving dust distribution. Our result will thus have wider validity, at the cost of a reduced predictivity/falsifiability. We will base our prescription on the work of \citet{Okuzumi2012}, \citet{Krijt2015}, \citet{Estrada2022} and \citet{Tazaki2021}.

Our goal is to improve the understanding of the effect and importance of including porosity into computations of the dust evolution in global models of protoplanetary disks, especially in the light of the importance of the fragmentation barrier and the bouncing barrier.

The paper is structured as follows. In Sect. \ref{sec:model_setup}, we describe the general model set-up. In Sect. \ref{sec:sc_models}, we apply a prescription for fractal growth using simple expressions for the cross sections and present the results of these models. Thereafter, we present models with more elaborate expressions for the cross sections in Sect. \ref{sec:ac_models}. Then, we discuss some aspects of our results in Sect. \ref{sec:discussion} and summarize our conclusions in Sect. \ref{sec:conclusions}.

\section{General model set-up}
\label{sec:model_setup}
We used the Python tool \texttt{DustPy} v.1.0.6  \citep{StammlerBirnstiel2022} to model the dust evolution. \texttt{DustPy} applies a 1-dimensional disk model and solves the advection-diffusion equation for gas and dust transport in the radial direction, as well as the Smoluchowski equation for dust growth.
\subsection{Disk model}
Our disk ranges from \(r_\text{in} \!=\! 1\ \mathrm{au}\) to \(r_\text{out} \!=\! 10^3\ \mathrm{au}\), with 100 logarithmically-spaced grid cells. The central star has a mass of \(M_\star \!=\! 1M_\odot\), a radius of \(R_\star \!=\! 2R_\odot\) and a temperature of \(T_\star \!=\! 5772\ \mathrm{K}\).

For the initial gas surface density, we apply the self-similar solution of \citet{Lynden-BellPringle1974}, which is given by
\begin{equation}
	\Sigma_\mathrm{g}(r) = \frac{M_\text{gas}}{2\pi r_\mathrm{c}^2} \left(\frac{r}{r_\mathrm{c}}\right)^{-1}\exp\left(-\frac{r}{r_\mathrm{c}}\right),
\end{equation}
where we chose the cut-off radius \(r_\mathrm{c} \!=\! 60\ \mathrm{au}\) and a gas mass of \(M_\mathrm{gas} \!=\! 10^{-2}M_\odot\).

For the temperature, it is assumed that the disk is passively irradiated at a constant irradiation angle of 0.05 rad, which yields
\begin{equation}
	T(r) = T_\star \left(\frac{0.05}{2}\right)^{1/4}\left(\frac{r}{R_\star}\right)^{-1/2}.
\end{equation}
In the direction vertical to the disk midplane, isothermal hydrostatic equilibrium is assumed \citep{StammlerBirnstiel2022}. The mean mass of gas particles is \(\overline{m} \!=\! 2.3m_\mathrm{p}\), where \(m_\mathrm{p}\) is the proton mass.

\subsection{Dust model}
The mass grid is logarithmic with 7 bins per decade. The grid ranges from \(m_\mathrm{min}=10^{-12}\ \mathrm{g}\) to \(m_\mathrm{max} \!=\! 10^5\ \mathrm{g}\) when bouncing is active, and to \(m_\mathrm{max} \!=\! 10^8\ \mathrm{g}\) when bouncing is inactive.

Initially, the dust-to-gas-ratio \(\Sigma_\mathrm{d} / \Sigma_\mathrm{g}\) is set to 0.01. For the initial size distribution, we adopted the Mathis-Rumple-Nordsieck (MRN) distribution \(n(a) \!\propto\! a^{-7/2}\) \citep{Mathis1977}, up to a maximum size of \(a_\mathrm{ini,max} \!=\! 10^{-4}\ \mathrm{cm}\). In the outer part of the disk, even such small grains can have drift time scales smaller than their growth time scales. These grains are removed from the initial dust distribution (for details, see \texttt{DustPy} documentation\footnote{\url{https://stammler.github.io/dustpy/}}).

We assumed a material density of \(\rho_\mathrm{s} \!=\! 1.67\ \mathrm{g\ cm^{-3}}\). Dust aggregates are a collection of small dust grains, so-called monomers. For a standard \texttt{DustPy} simulation, it is not necessary to determine the monomer size. However, since we included bouncing and fractal growth, we needed to determine the monomer size, which we set to \(a_\bullet \!=\! 10^{-4}\ \mathrm{cm}\). Although monomer grains are likely of irregular shape in reality \citep{Guettler2019}, we assumed them to be compact spheres for simplicity.

The vertical dust density profile is assumed to be Gaussian with the dust scale height \citep{Dubrulle1995}
\begin{equation}
h = H_\mathrm{P}\sqrt{\frac{\alpha}{\alpha + \mathrm{St}}},
\end{equation}
where \(H_\mathrm{P}\) is the pressure scale height of the gas, \(\alpha\) is the turbulent viscosity parameter introduced by \citet{ShakuraSunyaev1973}, and \(\mathrm{St}\) is the Stokes number, which is a measure for the coupling between dust and gas. It is given by \citep{StammlerBirnstiel2022}
\begin{equation}
    \label{eq:Stokes}
	\mathrm{St} = \begin{cases}
		\displaystyle \frac{\pi}{2}\frac{a\rho}{\Sigma_\mathrm{g}} \  & \text{for } a < \frac{9}{4}\lambda_\mathrm{mfp} \text{ (Epstein regime),}\\
		\displaystyle \frac{2\pi}{9}\frac{a^2\rho}{\lambda_\mathrm{mfp}\Sigma_\mathrm{g}} \  & \text{else} \text{ (Stokes regime),}\\
	\end{cases}
\end{equation}
where \(a\) is the size of the particle, \(\rho\) is the bulk density of the particle and \(\lambda_\mathrm{mfp}\) is the mean free path of the gas.

\subsection{Relative velocities}
\texttt{DustPy} considers Brownian motion, radial and azimuthal drift, vertical settling and turbulence as sources for relative velocities between dust particles. For the turbulent relative velocities, the model of \citet{OrmelCuzzi2007} is applied, which considers different turbulence regimes. The turbulent relative velocities steeply increase, when the stopping time \(t_\mathrm{s} = \mathrm{St}/\Omega_\mathrm{K}\) of the larger particle exceeds the Kolmogorov time \(t_\eta=\Omega_\mathrm{K}^{-1}\mathrm{Re}^{-1/2}\), which corresponds to the eddy turnover time of the smallest turbulent eddies. \(\mathrm{Re}=\nu_\mathrm{t}/\nu_\mathrm{mol}\) is the Reynolds number, which is defined as the ratio between turbulent viscosity \(\nu_\mathrm{t}=\alpha c_\mathrm{s}H_\mathrm{P}\) and molecular viscosity \(\nu_\mathrm{mol}=\sqrt{8/\pi}c_\mathrm{s}\lambda_\mathrm{mfp}/2\),  where \(c_\mathrm{s}\) is the sound speed\footnote{\texttt{DustPy} v.1.0.6 uses the adiabatic sound speed to compute the turbulent relative velocities, with an adiabatic coefficient of \(\gamma \!=\! 1.4\). In newer versions of \texttt{DustPy}, the isothermal sound speed is used, which is more appropriate. However, the deviation is only small (a factor of \(\sqrt{\gamma}\)).}.

The contributions of turbulence (for \(t_\mathrm{s}<t_\eta\)), radial drift, azimuthal drift and vertical settling to the relative velocities depend on the difference between the Stokes numbers of the collision partners. Therefore, for two particles with the same Stokes number, the only contribution to relative velocities is Brownian motion. This is in particular relevant for a fractal dimension of \(D\!=\!2\), because in this case, the Stokes number, which is proportional to the mass-to-area ratio \(m/A_\mathrm{aero}\), does not depend on mass if the aerodynamical cross sections are given by \(A_\mathrm{aero} = \pi a^2\). If more elaborate expressions are applied for the cross sections,  \(m/A_\mathrm{aero}\) approaches a constant value for large \(N\) if \(D\lesssim2\) \citep[e.g.,][]{Ormel2007,Okuzumi2011}. However, as pointed out by \citet{Okuzumi2011}, the dispersion of cross section area (from now on referred to as area dispersion) can not be neglected in this case, meaning that two particles of the same mass can have slightly different cross sections due to statistical variations.

We assumed that Stokes numbers follow a narrow log-normal distribution for each particles mass. We then averaged over the relative velocity equations and expressed them in terms of the mean Stokes numbers of the respective masses. This lead to the appearance of additional terms, which are non-zero, even if the mean Stokes numbers are the same. The detailed derivation and the resulting expressions are shown in Appendix \ref{sec:area_dispersion_derivation}.

The width of the dispersion is parametrized by the parameter \(\epsilon\). \citet{Okuzumi2011} measured the value of \(\epsilon\) in their numerical models of fractal aggregates and found \(\epsilon\approx0.1\), which we adopted.
To verify its importance, we also ran models without area dispersion, which are shown in Appendix \ref{sec:discussion_area_dispersion}.

\subsection{Collision model}
\label{sec:collision_model}

By default, \texttt{DustPy} has two possible collision outcomes: sticking and fragmentation. If the relative velocity \(v_\mathrm{rel}\) between two collision partners is below a threshold velocity \(v_\mathrm{f}\), then the collision results in perfect sticking. Otherwise, the collision results in fragmentation. \texttt{DustPy} further differentiates between full fragmentation for similar-sized aggregates, and erosion, if the mass ratio of the collision partners exceeds a critical mass ratio \(r_m\), which we set to \(r_m=10\). In a full fragmentation event, the fragment mass is the total mass of both collision partners, in an erosion event, the larger particle only loses a part of its mass, which we chose to be equal to the mass of the smaller particle. The fragment distribution is \(f(m)\mathrm{d}m\propto m^{-11/6}\mathrm{d}m\), which again corresponds to the MRN distribution.

In \texttt{DustPy}, the relative velocities are assumed to follow a Maxwell-Boltzmann distribution, where the single velocity computed for each pair of masses at each radial location corresponds to the root-mean-square velocity of the Maxwell-Boltzmann distribution. Thus, even if the root-mean-square velocity is larger than the fragmentation velocity \(v_\mathrm{f}\), there is a chance that the collision partners have a smaller velocity and the collision does not result in fragmentation. For details, see \citet{StammlerBirnstiel2022}.

\citet{DominikDullemond2024} extend the collision model to also take into account bouncing. They define the bouncing velocity as
\begin{equation}
	v_\mathrm{b}(m_1, m_2) = \min\left[ v_\mathrm{s}(m_1, m_2), v_\mathrm{f} \right],
	\label{eq:v_b}
\end{equation}
where \(m_1\) and \(m_2\) are the masses of the collision partners, respectively, and \(v_\mathrm{s}\) is the sticking velocity, which is given by \citep{DominikTielens1997,Guettler2010}
\begin{equation}
	\label{eq:v_s}
	v_\mathrm{s}(m_1, m_2) = \sqrt{5\pi a_\bullet F_\mathrm{roll}\frac{m_1 + m_2}{m_1 m_2}},
\end{equation}
with the rolling force \(F_\mathrm{roll}\). It is defined as the force that is necessary to roll a monomer over another monomer. Collisions with relative velocities between \(v_\mathrm{b}\) and \(v_\mathrm{f}\) result in bouncing, i.e., the masses of both collision partners stay the same. For details on the implementation of bouncing into \texttt{DustPy}, see \citet{DominikDullemond2024}.

Studies suggest that the filling factor also affects the collision outcomes \citep[e.g.,][]{Guettler2010,Wada2011,Gunkelmann2016,Arakawa2023,Oshiro2025}. However, there is still much uncertainty when it comes to the influence of the filling factor on the collision outcomes.

To keep things simple, we did not consider the effect of the filling factor on the collision outcome, but we investigated models with bouncing and models without bouncing. The limitations of our simple collision model are discussed in Sect. \ref{sec:collision_discussion}. Following \citet{DominikDullemond2024}, we set the fragmentation velocity to \(v_\mathrm{f} \!=\! 100\ \mathrm{cm\ s^{-1}}\) and the rolling force to \(F_\mathrm{roll} = 10^{-4}\ \mathrm{dyn}\).

\section{Fractal model}
\label{sec:sc_models}
In this work, we did not model fractal growth by evolving the volume filling factor numerically. Instead, we modeled fractal growth by using a fixed prescription for the filling factor as a function of mass, and treated the fractal dimension as a free parameter. This allowed us to investigate different degrees of fractality, i.e., different fractal dimensions. 

In this section, we treat the cross sections of fractal particles in a simplified way by approximating them with the cross sections of spherical particles (for details, see Sect. \ref{sec:sc_porosity_model}). This allows us to easily adjust existing estimates for the growth barriers to the case of fractal growth (Sect. \ref{sec:growth_barriers} and Appendix \ref{sec:appendix_barriers}), which helps us to better understand the numerical results. In Sect. \ref{sec:ac_models}, we apply a more elaborate, semi-analytic expression for the cross sections of fractal particles, and compare the results to those of the models with the simple expression.

\subsection{Porosity model}
\label{sec:sc_porosity_model}
The filling factor \(\phi\) of a dust aggregate is defined as the volume fraction that is filled with material. Hence, it can be written as the ratio \(\phi = \rho/\rho_\mathrm{s}\) between the bulk density \(\rho\) of the aggregate and the material density \(\rho_\mathrm{s}\) of the monomer grains \citep[e.g.,][]{Kataoka2013a}.

In the first phase of dust coagulation, the impact energies are too small to lead to significant compaction of the aggregates. Therefore, the particles become more and more porous, as they grow \citep{DominikTielens1997}, and the filling factor decreases. This can be described by the fractal dimension \(D\), which is defined by the relation 
\begin{equation}
	m\propto a^D.
\end{equation}
The fractal dimension is \(D\!=\!3\) for non-fractal growth and \(D\!<\!3\) for fractal growth. 

The definition of radius of a fractal particle is ambiguous. Often the radius of gyration is used \citep[e.g.,][]{Wada2007,Wada2008,Okuzumi2009b}, which is given by
\begin{equation}
	a_\mathrm{g} = \sqrt{\frac{1}{N}\sum_{k=1}^{N}\left(\mathbf{x}_k - \mathbf{x}_\mathrm{s}\right)^2},
\end{equation}
where \(N\) is the number of monomers, \(\mathbf{x}_k\) (\(k=1,2,...,N\)) are their positions and \(\mathbf{x}_\mathrm{s}\) is the center of mass. 
The gyration radius follows the relation \citep{Tazaki2021}
\begin{equation}
	\label{eq:gyration_radius}
	a_\mathrm{g} = a_\bullet \left(\frac{m}{k_0 m_\bullet}\right)^{1/D},
\end{equation}
where \(m_\bullet\) is the monomer mass and \(k_0\) is the fractal pre-factor, which depends on the fractal dimension. Equation \eqref{eq:gyration_radius} does not hold for small \(m\) if \(k_0\neq1\), since it does not recover the monomer radius for \(m=m_\bullet\). 

For the simple prescription of the filling factor in this section, we omit the fractal pre-factor in Eq. \eqref{eq:gyration_radius} and define the particle radius in terms of mass and fractal dimension as
\begin{equation}
	a = a_\bullet \left(\frac{m}{m_\bullet}\right)^{1/D}.
\end{equation}
At some point, collision energies become large enough to lead to restructuring and compaction of the aggregates \citep{DominikTielens1997,Guettler2010}. 

\citet{Michoulier2024} modeled the evolution of porosity and included collisional compaction. In their models, fractal growth proceeds also after collisional compaction sets in, but with a larger fractal dimension than before.
Other studies find that fractal growth terminates once collision energies become large enough to lead to compaction. The filling factor then stays approximately constant \citep{Okuzumi2012,Krijt2015,Estrada2022}. 

We accounted for this by introducing a minimum filling factor \(\phi_\mathrm{min}\). When particles reach this minimum filling factor, growth proceeds non-fractally and the filling factor stays constant. In reality, the minimum filling factor is a function of radial distance \citep{Okuzumi2012} but for simplicity, we assumed it to be constant throughout the disk. Furthermore, very large aggregates can be also compressed by gas ram pressure and self-gravity \citep{Kataoka2013b,Estrada2022,Michoulier2024}. Since the focus of this work is on the influence of the fractal dimension, we did not consider compaction by gas ram pressure and self-gravity.

The mass domain extends to smaller masses than the monomer mass. For masses \(m < m_\bullet\), we assumed \(\phi=1\). Therefore, the full expression for the particle radius as a function of mass is given by
\begin{equation}
	\label{eq:sc_radius}
	a(m) = \begin{cases}
		\displaystyle a_\bullet \left(\frac{m}{m_\bullet}\right)^{1/3}\  & \text{for } m < m_\bullet,\\
		\displaystyle a_\bullet \left(\frac{m}{m_\bullet}\right)^{1/D}\  & \text{for } m_\bullet \leq m \leq m_\bullet \phi_\mathrm{min}^{D/(D-3)},\\
		\displaystyle a_\bullet \phi_\mathrm{min}^{-1/3}\left(\frac{m}{m_\bullet}\right)^{1/3}\  & \text{for }  m_\bullet \phi_\mathrm{min}^{D/(D-3)} < m,
	\end{cases}
\end{equation}
and the full expression for the filling factor is given by
\begin{equation}
	\label{eq:sc_phi}
	\phi(m) = \begin{cases}
		\displaystyle 1\  & \text{for } m < m_\bullet,\\
		\displaystyle \max\left[\left(\frac{m}{m_\bullet}\right)^{1-3/D}, \phi_\mathrm{min}\right]\  & \text{else}.
	\end{cases}
\end{equation}

\begin{figure}
   \centering
   \includegraphics[width=\hsize]{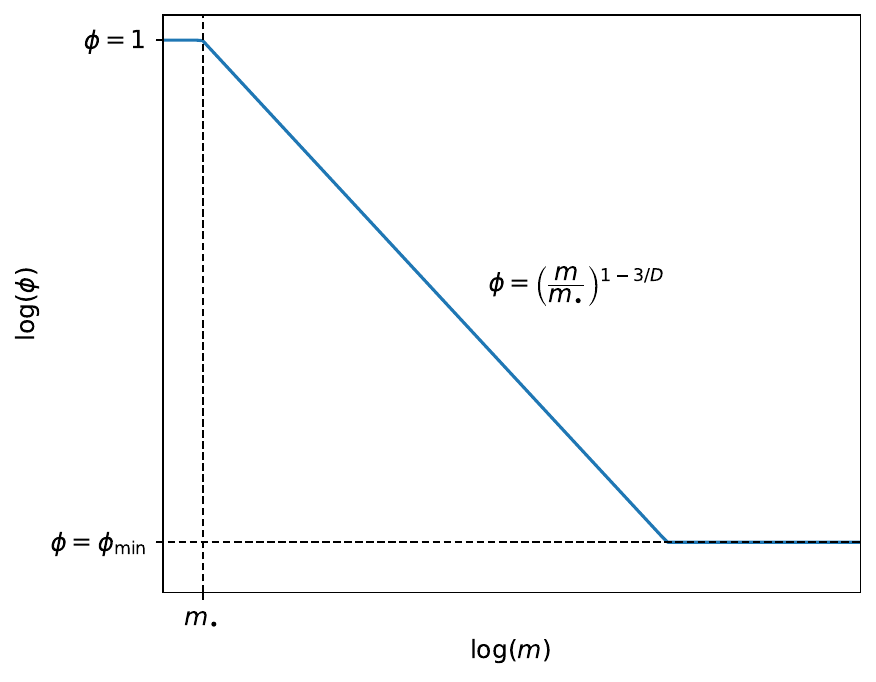}
      \caption{Filling factor \(\phi\) as a function of mass for the SC models.}
         \label{fig:sc_phi}
\end{figure}

The collision cross section between particles is given by 
\begin{equation}
	\label{eq:A_coll}
	A_\mathrm{coll} = \pi\left(a_1 + a_2\right)^2,
\end{equation}
where \(a_1\) and \(a_2\) are the radii of the collision partners, respectively. For the aerodynamical cross section, we applied the expression
\begin{equation}
	\label{eq:simple_A}
	A_\mathrm{aero} = \pi a^2,
\end{equation}
which corresponds to the cross section of a sphere of radius \(a\). This is of course a simplification, because in reality, fractal particles are not spherical and their cross sections take a more complicated form. We therefore also investigated models with more elaborate cross sections, which are presented in Sect. \ref{sec:ac_models}.

\begin{figure}
   \centering
   \includegraphics[width=\hsize]{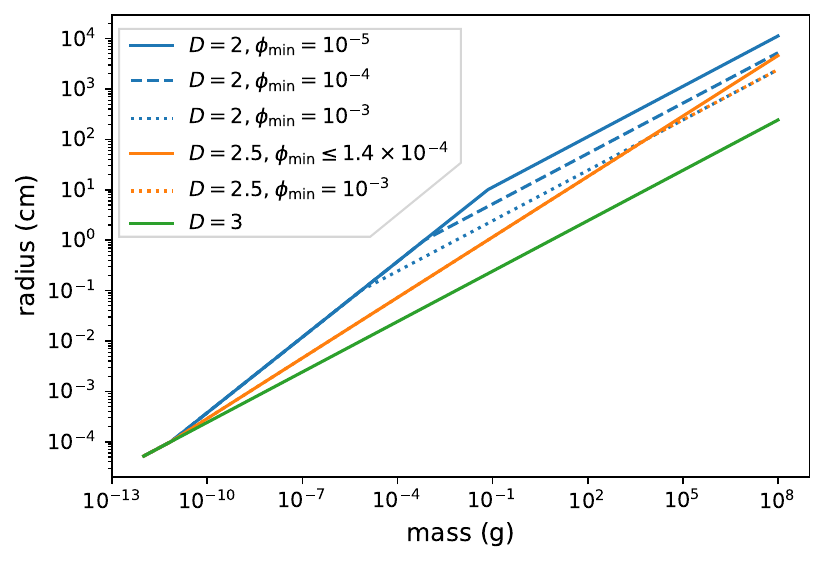}
      \caption{Mass-radius relation for the SC models.}
         \label{fig:sc_m_a_relations}
\end{figure}

Using the expression of Eq. \eqref{eq:simple_A} has the consequence that the fractal dimension can only take values between \(2\leq D \leq 3\). Values \(D\!<\!2\) would imply that the aerodynamical cross section of an aggregate is larger then the sum of the aerodynamical cross sections of all its monomers, which is unphysical. We will refer to the models that apply the above porosity prescription as simple cross section (SC) models. Figure \ref{fig:sc_phi} shows the filling factor as a function of mass for the SC models as given by Eq. \eqref{eq:sc_phi}. The mass-radius relations of the SC models are shown in Fig. \ref{fig:sc_m_a_relations}.

\begin{table}
	\centering
	\caption{Properties of the SC models}
	\label{tab:sc_models}
	\begin{tabular}{c|lllll}
\hline\hline
&Model name  & \(D\) & \(\phi_\mathrm{min}\) &\(\alpha\)	& bo.?\\

\hline
\multirow{6}{*}{\rotatebox{90}{\shortstack{Vary \(D\)}}}

&\model{SC\_D3\_a4}     				& 3		  & -	 		& \(10^{-4}\) & yes \\
&\model{SC\_D25\_fm5\_a4} 			& 2.5	  & \(10^{-5}\)	& \(10^{-4}\) & yes \\
&\model{SC\_D2\_fm5\_a4}				& 2		  & \(10^{-5}\)	& \(10^{-4}\) & yes \\

\cline{2-6}

&\model{SC\_D3\_a4\_nobo}			& 3		  & -		 		& \(10^{-4}\) & no \\
&\model{SC\_D25\_fm5\_a4\_nobo}		& 2.5	  & \(10^{-5}\)	& \(10^{-4}\) & no \\
&\model{SC\_D2\_fm5\_a4\_nobo}		& 2		  & \(10^{-5}\)	& \(10^{-4}\) & no \\

\hline
\multirow{4}{*}{\rotatebox{90}{\shortstack{Vary \(\phi_\mathrm{min}\)}}}

&\model{SC\_D2\_fm3\_a4} 			& 2		  & \(10^{-3}\)	& \(10^{-4}\)	& yes \\
&\model{SC\_D2\_fm4\_a4} 			& 2		  & \(10^{-4}\)	& \(10^{-4}\)	& yes \\

\cline{2-6}

&\model{SC\_D2\_fm3\_a4\_nobo} 		& 2		  & \(10^{-3}\)	& \(10^{-4}\)	& no \\
&\model{SC\_D2\_fm4\_a4\_nobo} 		& 2		  & \(10^{-4}\)	& \(10^{-4}\)	& no \\

\hline
\multirow{12}{*}{\rotatebox{90}{\shortstack{Vary \(\alpha\)}}}

&\model{SC\_D3\_a6} 					& 3		  & -			& \(10^{-6}\)	& yes\\
&\model{SC\_D3\_a3} 					& 3		  & -			& \(10^{-3}\)	& yes\\
&\model{SC\_D25\_fm5\_a6} 			& 2.5	  & \(10^{-5}\)	& \(10^{-6}\)	& yes\\
&\model{SC\_D25\_fm5\_a3} 			& 2.5	  & \(10^{-5}\)	& \(10^{-3}\)	& yes \\
&\model{SC\_D2\_fm5\_a6} 			& 2		  & \(10^{-5}\)	& \(10^{-6}\)	& yes \\
&\model{SC\_D2\_fm5\_a3} 			& 2		  & \(10^{-5}\)	& \(10^{-3}\)	& yes \\

\cline{2-6}

&\model{SC\_D3\_a6\_nobo} 			& 3		  & -			& \(10^{-6}\)	& no	\\
&\model{SC\_D3\_a3\_nobo} 			& 3		  & -			& \(10^{-3}\)	& no \\
&\model{SC\_D25\_fm5\_a6\_nobo} 		& 2.5	  & \(10^{-5}\)	& \(10^{-6}\)   	& no \\
&\model{SC\_D25\_fm5\_a3\_nobo} 		& 2.5	  & \(10^{-5}\)	& \(10^{-3}\)	& no \\
&\model{SC\_D2\_fm5\_a6\_nobo} 		& 2		  & \(10^{-5}\)	& \(10^{-6}\)	& no \\
&\model{SC\_D2\_fm5\_a3\_nobo} 		& 2		  & \(10^{-5}\)	& \(10^{-3}\)	& no \\
\hline
\end{tabular}
	\tablefoot{The last column shows whether bouncing is included or not. The \model{SC} in the model name stands for the simple cross sections. The number following \model{D} stands for the fractal dimension, where the decimal point is omitted. The number following \model{fm} is the negative logarithm of \(\phi\). The number following \model{a} is the negative logarithm of \(\alpha\). If the name ends with \model{nobo}, the model does not include bouncing.}
\end{table}

\subsection{Growth barriers}
\label{sec:growth_barriers}
The dust distribution is shaped by several growth barriers. The drift barrier is related to the fact that at some Stokes number, the particles drift inward rapidly enough to get lost in the evaporation zone, before they can grow to larger sizes \citep[e.g.,][]{Weidenschilling1977,Brauer2008}. 
The fragmentation \citep[e.g.,][]{Birnstiel2012} and the bouncing \citep[e.g.,][]{BlumWurm2008,Zsom2010,DominikDullemond2024} barrier are related to the fact that collisions result in bouncing or fragmentation, once the collision energies exceed the corresponding threshold (see Sect. \ref{sec:collision_model}). 

Following \citet{Birnstiel2012}, we compare the drift time scale \(\tau_\mathrm{drift}=r/|v_\mathrm{d}|\), where \(r\) is the radial distance to the central star and \(v_\mathrm{d}\) is the drift velocity, to the growth time scale \(\tau_\mathrm{grow}=a/\dot{a}\), where \(\dot{a}\) is the change of size per unit time of the particle. This yields the maximum Stokes number that grains can reach before drifting inward faster than growing further \citep{Birnstiel2012}:
\begin{equation}
	\label{eq:St_d}
	\mathrm{St_d} = \frac{\Sigma_\mathrm{d}}{\Sigma_\mathrm{g}}\frac{v_\mathrm{K}^2}{c_\mathrm{s}^2}\left|\frac{\mathrm{d}\ln P}{\mathrm{d}\ln r}\right|^{-1},
\end{equation}
where \(P\) is the gas pressure and \(v_\mathrm{K}=r\Omega_\mathrm{K}\) is the Kepler velocity, with the Kepler frequency \(\Omega_\mathrm{K}\). Eq. \eqref{eq:St_d} shows that \(\mathrm{St_d}\) does not depend on particle properties.

The Stokes number corresponding to the fragmentation barrier can be derived by comparing the relative velocities to the fragmentation velocity and is given by \citep{Birnstiel2009}
\begin{equation}
	\label{eq:St_f}
	\mathrm{St_f} = \frac{1}{3}\frac{v_\mathrm{f}^2}{\alpha c_\mathrm{s}^2},
\end{equation}
where it is assumed that turbulence dominates the relative velocities and \(v_\mathrm{rel}\approx 3\sqrt{\alpha\mathrm{St}}c_\mathrm{s}\). 
The Stokes number corresponding to the bouncing barrier is given by \citep{DominikDullemond2024}
\begin{equation}
	\label{eq:St_b}
	\mathrm{St_{b}} = \max\left[\frac{1}{3}\frac{v_\mathrm{b}^2}{\alpha c_\mathrm{s}^2}, \mathrm{St}_\eta\right],
\end{equation}
where \(\mathrm{St}_\eta=\mathrm{Re}^{-1/2}\) is the Stokes number, at which the particles stopping time \(t_\mathrm{s}\) exceeds the Kolmogorov time \(t_\eta\). We introduced it following \citet{DominikDullemond2024} to account for the fact that the contribution of turbulence to the relative velocities is much smaller for \(t_\mathrm{s} < t_\eta\) than for \(t_\mathrm{s} > t_\eta\). In contrast to \(v_\mathrm{f}\), the bouncing velocity is a function of mass. Thus, while \(\mathrm{St_f}\) does not depend on particle properties (except for the value of \(v_\mathrm{f}\)), \(\mathrm{St_b}\) is not independent of particle properties, since the relation between \(\mathrm{St}\) and mass depends on properties such as the fractal dimension or filling factor. 

Eqs. \eqref{eq:St_d}, \eqref{eq:St_f} and \eqref{eq:St_b} can be solved for the mass to yield the respective growth barriers in terms of mass. This is shown in Appendix \ref{sec:appendix_barriers}. The estimates predict that the dust distribution of models with smaller fractal dimensions reach larger masses due to their smaller Stokes numbers, which lead to smaller relative velocities.

The smallest particles of a bouncing-limited dust distribution collide mainly with larger particles. The minimum energy required for collisions to result in bouncing is set by the adopted bouncing threshold velocity (see Eqs. \eqref{eq:v_b} and \eqref{eq:v_s}). From Eq. \eqref{eq:v_s}, it can be deduced that the collision energy involved in collisions with larger particles is typically larger than that involved in collisions with equal-sized or smaller particles. Therefore, the smallest particles of the mass distribution reach the bouncing barrier already at smaller masses. We exploit this fact to derive an estimate of the lower edge of the mass distribution, which we will refer to as the "lower bouncing barrier", as opposed to the usual bouncing barrier, which corresponds to the upper edge of the mass distribution. We derive the lower bouncing barrier in Appendix \ref{sec:appendix_barriers}.

\subsection{Influence of the fractal dimension}
\label{sec:sc_varyD}
\begin{figure*}
   \centering
   \includegraphics[width=\hsize]{figures/sc_sigma_d_varyD.pdf}
      \caption{Dust distributions of the SC models after \(10^6\) years with bouncing (upper row) and without bouncing (lower row) for different fractal dimensions. The green, blue, red and yellow lines show the analytical estimates of the drift, fragmentation, bouncing, and lower bouncing barrier, respectively. The white dashed lines indicate the transition from the fractal to the non-fractal growth regime. For \(D=2.5\), this transition lies outside the shown mass range.}
         \label{fig:sc_sigma_d_varyD}
\end{figure*}
\begin{figure*}
   \centering
   \includegraphics[width=\hsize]{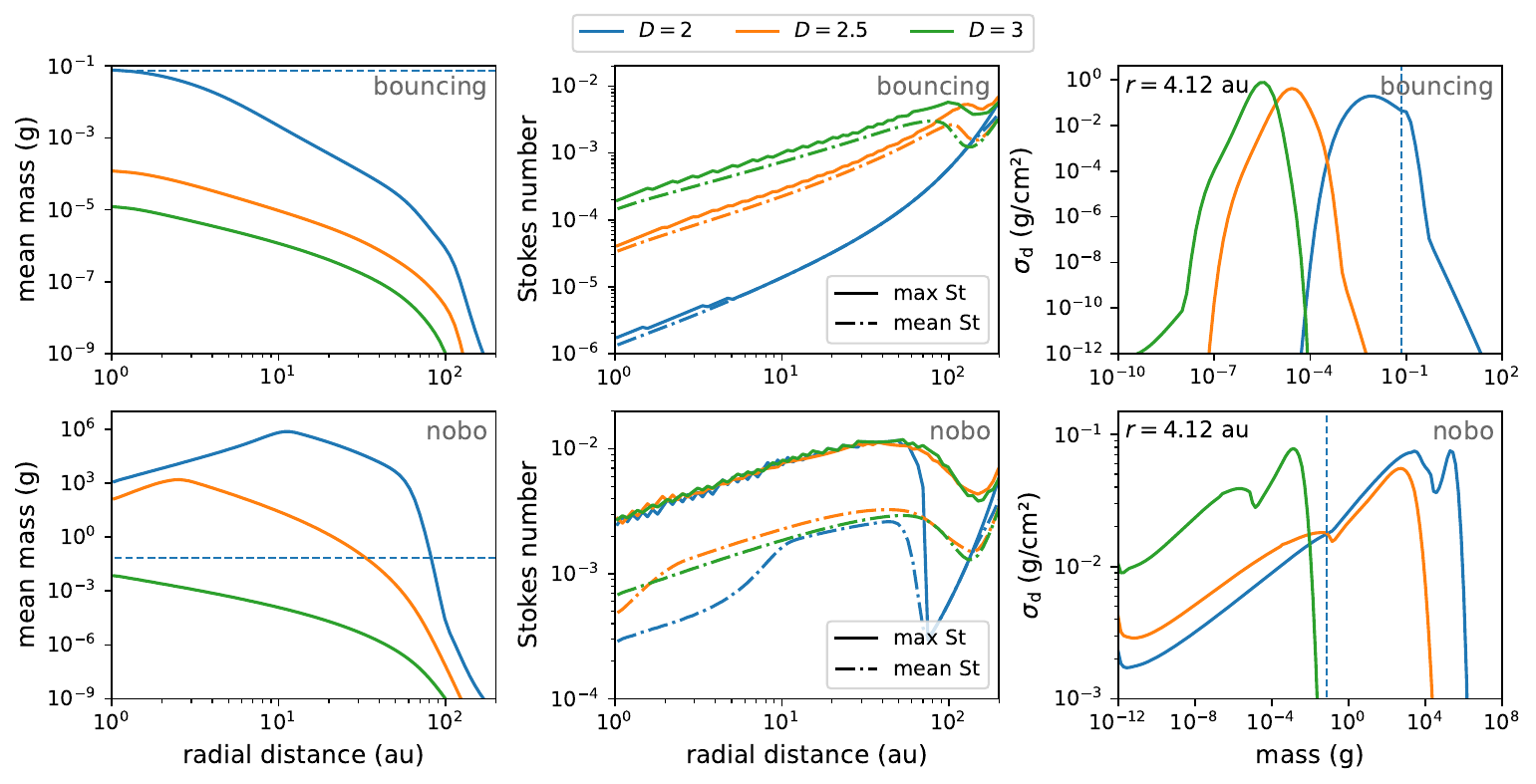}
      \caption{The mean mass (left), the maximum and mean Stokes numbers (middle), and a radial slice at \(r=4.12\) au (right) of the mass distribution after \(10^6\) years for different fractal dimensions of the models with bouncing (upper row) and without bouncing (lower row). The blue dashed lines in the right panels indicate the transition from fractal to non-fractal growth for the \(D\!=\!2\) models. For \(D=2.5\), this transition lies outside the shown mass range.}
         \label{fig:sc_mean_width_dist_varyD}
\end{figure*}

To investigate the influence of the fractal dimensions on the dust evolution, we ran models with highly fractal (\(D\!=\!2\)), moderately fractal (\(D\!=\!2.5\)), and non-fractal (\(D\!=\!3\)) growth. For these fiducial models, we chose \(\phi_\mathrm{min} \!=\! 10^{-5}\), roughly corresponding to the value found by previous studies \citep{Okuzumi2012,Krijt2015,Estrada2022}, and \(\alpha \!=\! 10^{-4}\), but also investigated different minimum filling factors (Sect. \ref{sec:sc_varyFillmin}) and turbulent viscosity parameters (Sect. \ref{sec:sc_varyAlpha}). For each fractal dimension, we ran a model with bouncing and one without bouncing. The properties of the models are shown in Table \ref{tab:sc_models}. The models were run for \(10^6\) years.

\subsubsection{Dust distribution}

The dust distributions \(\sigma_\mathrm{d}\) after \(10^6\) years are shown in Fig. \ref{fig:sc_sigma_d_varyD}. For the definition of \(\sigma_\mathrm{d}\), see Appendix \ref{sec:appendix_code_quantities}. As predicted by the analytical estimates, the dust distributions reach larger masses for smaller fractal dimensions (see also left panels of Fig. \ref{fig:sc_mean_width_dist_varyD}). 

It also becomes clear from Fig. \ref{fig:sc_sigma_d_varyD} that the analytical estimate overestimates the mass of the bouncing barrier for \(D\!<\!3\). For the analytical estimate it was assumed that the bouncing velocity is only exceeded when the larger particle leaves the \(t_\mathrm{s}<t_\eta\) regime. However, fractal particles can have rather large masses and still be in the \(t_\mathrm{s}<t_\eta\) regime. Since the bouncing velocity decreases with mass, it is exceeded already in the \(t_\mathrm{s}<t_\eta\) regime (if the mass ratio is not too large) despite the comparatively small relative velocities.
It should be noted that also for \(D \!=\! 3\) the bouncing velocity is already exceeded in the \(t_\mathrm{s}<t_\eta\) regime for collisions between similar masses. This suggests that collisions between particles with a large difference in mass  play a role, such that the particles can still grow further. The agreement of the estimate with the upper edge of the mass distribution in the numerical results for \(D \!=\! 3\) might therefore be a coincidence. Since the assumption of similar-sized collisions is not valid, we have to consider both masses for the bouncing velocity. Furthermore, in the \(t_\mathrm{s}<t_\eta\) regime the relative velocities depend on both Stokes numbers. Therefore, it is hard to come up with a good estimate for the bouncing barrier. In contrast, the estimate for the lower bouncing barrier works reasonably well.

For the models without bouncing, the particles can grow up to the fragmentation barrier. Since the fragmentation barrier lies outside the \(t_\mathrm{s}<t_\eta\) regime for all fractal dimensions, the estimate of the fragmentation barrier works well.

For \(D \!=\! 2\) and \(D \!=\! 2.5\), there is a kink in the fragmentation barrier at \(r\sim\!10\) au and \(r\sim4\) au, respectively. The reason is that the largest particles reach the Stokes drag regime in the inner parts of the disk. In the Stokes regime, the difference between the fragmentation barriers of the \(D \!=\! 2\) and \(D \!=\! 2.5\) model and the difference in mean mass is much smaller than in the Epstein regime, because the Stokes number, which is the relevant quantity determining the fragmentation barrier, depends more weakly on the filling factor in the Stokes regime than in the Epstein regime (for a fixed mass, \(\mathrm{St^{Ep.}}\propto \phi^{2/3}\) and \(\mathrm{St^{St.}}\propto\phi^{1/3}\), which can be derived by exploiting \(\rho\propto\phi\) and \(a\propto(m/\phi)^{1/3}\propto\phi^{-1/3}\)).

The qualitative shape of the fragmentation-limited mass distributions is similar for all fractal dimensions (see lower right panel of Fig. \ref{fig:sc_mean_width_dist_varyD}), approximately following a powerlaw.
\subsubsection{Stokes number}
As can be seen in the middle panels of Fig. \ref{fig:sc_mean_width_dist_varyD}, the maximum Stokes number that is reached by a fragmentation-limited dust distribution does not depend on the fractal dimension. Since \(\mathrm{St_f}\) does not depend on particle properties (see Eq. \eqref{eq:St_f}), the fact that larger masses are reached for smaller fractal dimensions can be understood as an effect of the Stokes number dependence on mass and fractal dimension. For smaller fractal dimensions, larger masses are necessary for the Stokes number to exceed \(\mathrm{St_f}\).

In the bouncing-limited case, the mean and maximum Stokes numbers decrease with decreasing fractal dimension. In contrast to \(\mathrm{St_f}\), \(\mathrm{St_b}\) depends on mass via the bouncing velocity (see Eq. \eqref{eq:St_b}). Models with smaller fractal dimensions can reach larger masses, because their Stokes numbers still stay smaller. However, the bouncing velocity decreases due to the larger masses, therefore, it is exceeded already at smaller Stokes numbers compared to models with larger fractal dimensions.

\subsubsection{Time evolution}
 Since the growth rate of dust particles depends on their collision cross section and their relative velocity, which both depend on the fractal dimension, we expect that the time evolution of the dust distribution is influenced by the fractal dimension. 
 
The growth rate is given by 
\begin{equation}
    \label{eq:growth_rate}
    \frac{\dot m}{m} = \int_{-\infty}^{\infty} \frac{R(m,m^\prime)}{m}\sigma_\mathrm{d}(m^\prime)\mathrm{d}\log m^\prime,
\end{equation}
with
\begin{equation}
    \frac{R(m,m^\prime)}{m} = \frac{A_\mathrm{coll}\left(m,m^\prime\right)v_\mathrm{rel}\left(m,m^\prime\right)}{m\sqrt{2\pi}h(m^\prime)}.
\end{equation}
Although the actual growth rate is given by Eq. \eqref{eq:growth_rate}, we will refer to \(R/m\) as the growth rate in the following consideration.

We consider the case, where Stokes numbers are in the Epstein regime, \(\mathrm{St}\ll\alpha\) and \(\mathrm{St}<\mathrm{St}_\eta = \mathrm{Re}^{-1/2}\). With the first condition, \(A_\mathrm{coll}/m\propto \mathrm{St}^{-1}\), where we assumed \(A_\mathrm{coll}\sim A_\mathrm{aero}\). With the second condition, we can approximate \(h\approx H_\mathrm{P}\).
With the last condition, the relative velocities, which are typically dominated by vertical settling and turbulence, can be written as 
\(
v_\mathrm{rel} \propto \sqrt{\epsilon^2 \mathrm{St}^2 + \left(\mathrm{St} - \mathrm{St}^\prime\right)^2}
\) (see Eq. \eqref{eq:v_vert_low_alpha} and Eqs. \eqref{eq:turb1a} and \eqref{eq:turb2}). Thus, the following relation approximately holds:
\begin{equation}
	\label{eq:R}
	\frac{R}{m} \propto \sqrt{\epsilon^2 + \frac{\left(\mathrm{St} - \mathrm{St}^\prime\right)^2}{\mathrm{St}^2}}
\end{equation}
For smaller \(D\), the term \(\mathrm{St}-\mathrm{St}^\prime\) is typically smaller, therefore, we expect the fractal particles to grow slower than the non-fractal particles, at least in the early phase of dust growth, where the above conditions are met.

\begin{figure}
   \centering
   \includegraphics[width=\hsize]{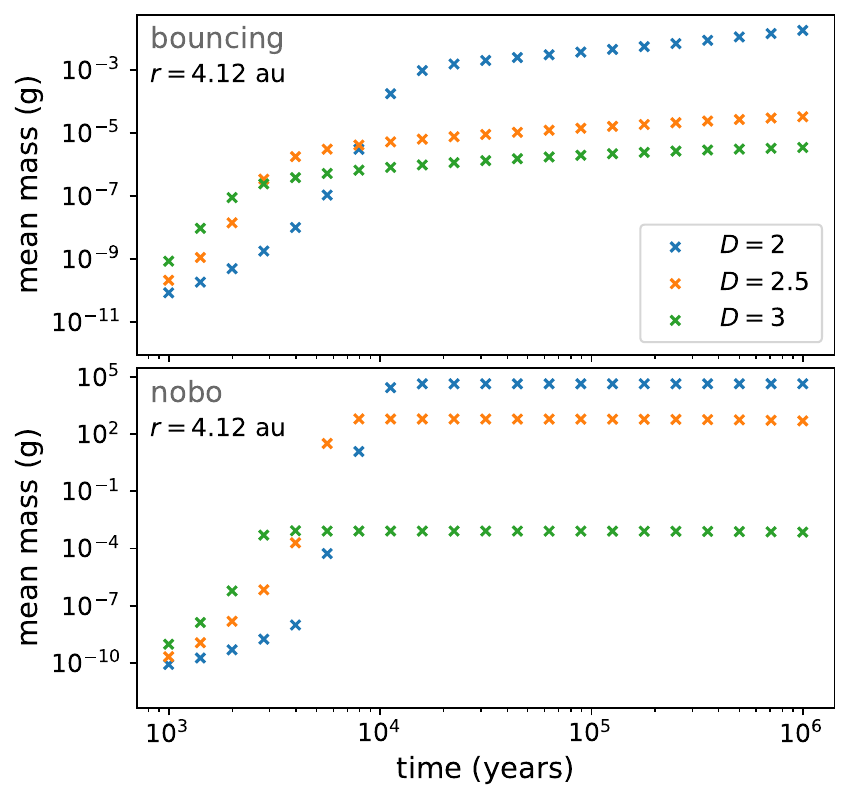}
      \caption{Time evolution of the mean mass for the models with bouncing (top) and without bouncing (bottom) at \(r=4.1\) au.}
         \label{fig:sc_m_t}
\end{figure}

When looking at the time evolution of the mean mass, shown in Fig. \ref{fig:sc_m_t}, growth is indeed slower for the fractal particles than the non-fractal particles, as long as the particles are in the sticking regime, i.e., before the growth barriers are reached. Interestingly, this is an apparent contradiction to previous studies that found that fractal growth is faster than non-fractal growth. In Sect. \ref{sec:discussion_time_evolution}, we discuss how this discrepancy could be explained.

While growth is slowed down drastically but not fully stopped at the bouncing barrier, the fragmentation barrier stops growth completely. Due to the Maxwell-Boltzmann distribution of the relative velocities, there is a non-zero probability that the collision velocity is low enough for the collision to result in sticking, even if the mean relative velocity is above the threshold velocity for bouncing or fragmentation, respectively. At the bouncing barrier, larger particles that are produced in these sticking collisions stay large, even if they undergo many bouncing collisions afterwards. In contrast, larger particles that are produced at the fragmentation barrier get destroyed soon afterwards due to fragmentation \citep{DominikDullemond2024}. 

\subsection{Variation of the minimum filling factor}
\label{sec:sc_varyFillmin}
The minimum filling factor determines the bulk density of the particles once they reach the non-fractal growth regime. Thus, we expect the dust distribution to depend on \(\phi_\mathrm{min}\) when the upper edge lies in the non-fractal growth regime.

Therefore, we varied the minimum filling factor. Since \(\phi_\mathrm{min} \!=\! 10^{-5}\) is already a rather low value, we investigated larger values \mbox{\(\phi_\mathrm{min} \!=\! 10^{-4}\)} and \mbox{\(\phi_\mathrm{min}  \!=\! 10^{-3}\)}. For \(D \!=\! 2.5\), even the comparatively large value \mbox{\(\phi_\mathrm{min}  \!=\! 10^{-3}\)} is only reached at \(m\approx 7\times10^{3}\mathrm{\ g}\), which is just below the fragmentation barrier. Thus, we did not expect a considerable influence of \(\phi_\mathrm{min}\) on the results of the \(D \!=\! 2.5\) models and limited our investigation to the \(D \!=\! 2\) models. 

\begin{figure}
   \centering
   \includegraphics[width=\hsize]{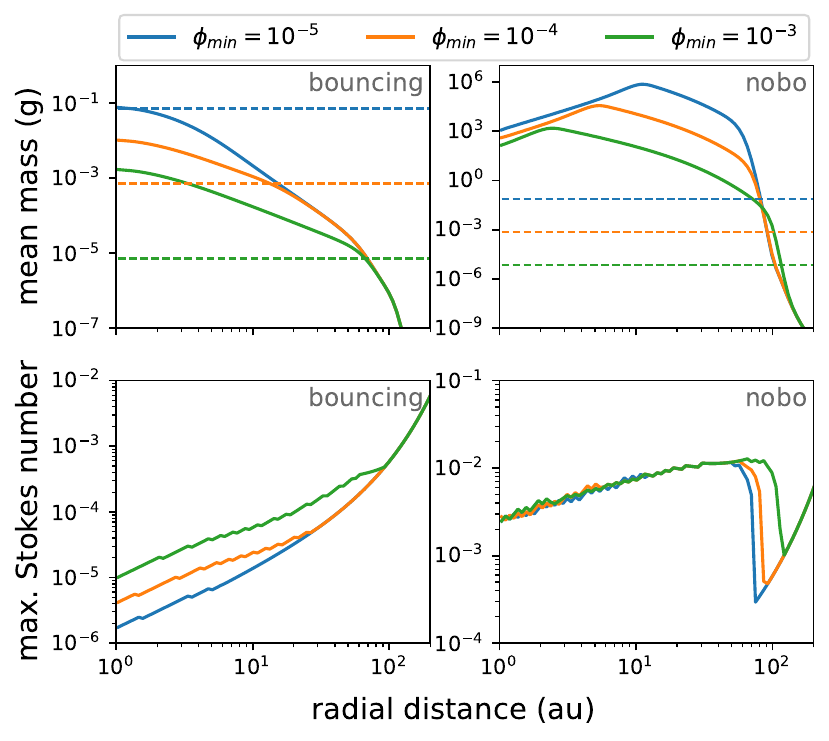}
      \caption{Variation of the minimum filling factor for the \(D=2\) models. Upper row: Mean mass. Lower row: Maximum Stokes number. Left: Models with bouncing. Right: Models without bouncing.}
         \label{fig:sc_mean_varyFillMin}
\end{figure}

The mean mass and the maximum Stokes number are shown in Fig. \ref{fig:sc_mean_varyFillMin}. Increasing the minimum filling factor shifts the transition to the non-fractal growth regime towards smaller masses. In the outer disk regions, the mean mass is still within the fractal growth regime for the models with bouncing, therefore, there is no significant difference in mean mass. In the inner parts of the disk, when the mean mass is in the non-fractal growth regime, the models with larger \(\phi_\mathrm{min}\) have smaller mean masses.  In the case without bouncing, all models reach the non-fractal growth regime in the whole disk. As for the models with bouncing, the mean mass is smaller for the models with larger \(\phi_\mathrm{min}\). The lower right panel of Fig. \ref{fig:sc_mean_varyFillMin} shows that the fragmentation barrier in terms of Stokes number does not depend on \(\phi_\mathrm{min}\). Thus, the difference in mean mass can again be understood as the consequence of the Stokes number dependence on mass and filling factor.  The maximum Stokes number reached by the models with bouncing decreases with increasing \(\phi_\mathrm{min}\). As in Sect. \ref{sec:sc_varyD}, this can be attributed to the mass-dependence of the bouncing velocity.

\subsection{Variation of the turbulent viscosity parameter}
\label{sec:sc_varyAlpha}
Many relevant processes and quantities depend on the turbulent viscosity parameter \(\alpha\). In particular, the vertical settling and the turbulent contributions to relative velocities depend on \(\alpha\). The influence of the viscosity parameter on the dust evolution is investigated in several studies \citep[e.g.,][]{Brauer2008,Estrada2022,DominikDullemond2024}. Here, we want to focus our investigation on how the reaction to a change of the turbulent viscosity parameter \(\alpha\) depends on the fractal dimension.

We therefore ran high turbulence models with \(\alpha=10^{-3}\), and low turbulence models with \(\alpha=10^{-6}\), each for different fractal dimensions. Note that model \model{SC\_D2\_fm5\_a3\_nobo} had a glitch at about \(t\sim 3.5\times10^5\) years, where the dust surface density dropped to zero in the inner part of the disk. However, the dust distribution reassembled again and the dust distribution at the end of the simulation after \(10^6\) years looks reasonable.

\begin{figure*}
   \centering
   \includegraphics[width=\hsize]{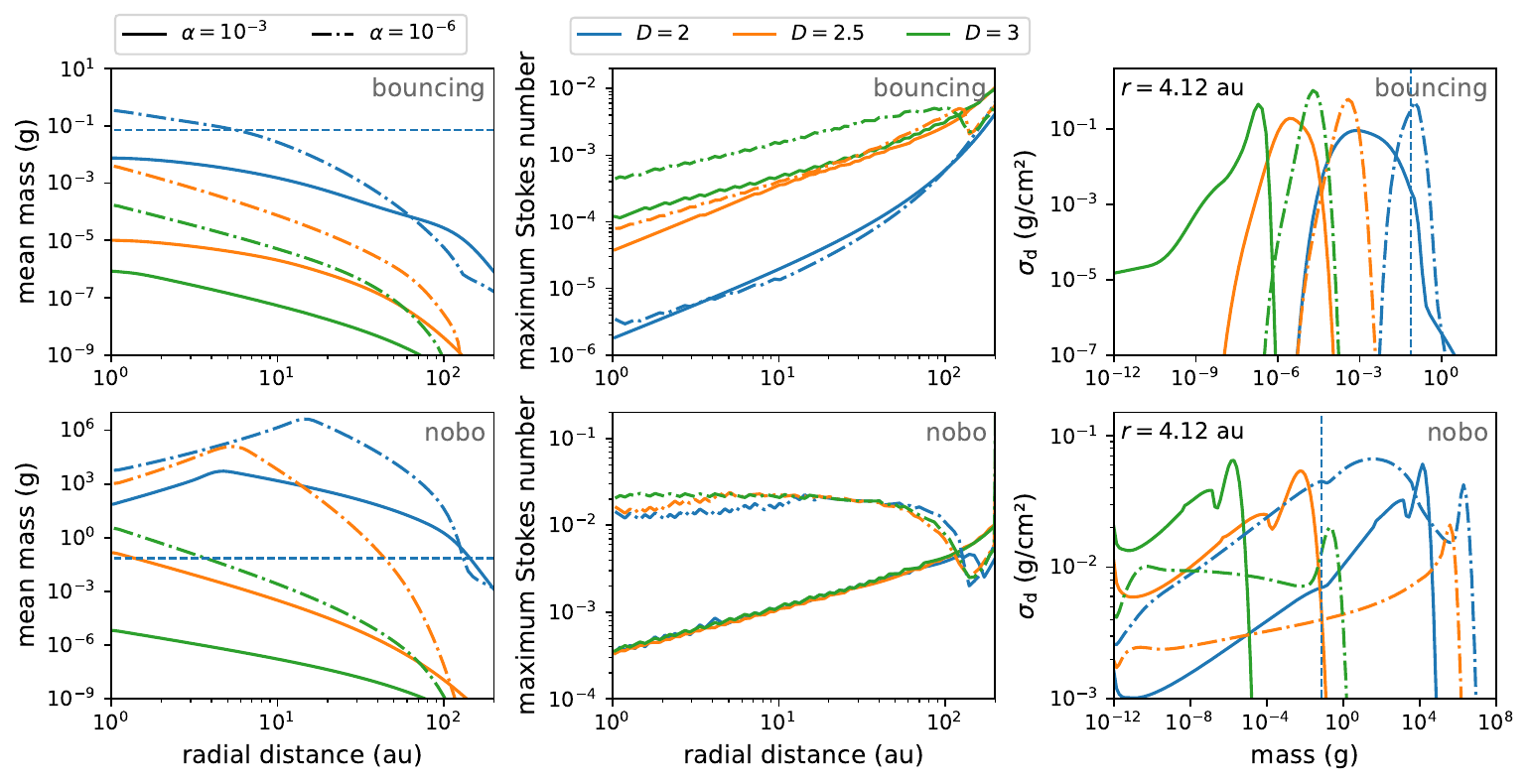}
      \caption{Mean mass (left), maximum Stokes number (middle) and a radial slice at \(r=4.1\) au of the dust distribution for high (\(\alpha=10^{-3}\); solid lines) and low turbulence (\(\alpha=10^{-6}\); dash-dotted lines). The blue dashed lines in the left panels indicate the transition from the fractal to the non-fractal growth regime. For \(D=2.5\), this transition lies outside the shown mass range.}
         \label{fig:sc_mean_width_dist_varyAlpha}
\end{figure*}

For the mean mass, the general behavior is the same for all fractal dimensions.
Since the relative velocities increase with the degree of turbulence, we expect the growth barriers shifted towards smaller masses for larger \(\alpha\), which is indeed the case for all fractal dimensions (see left panels of Fig. \ref{fig:sc_mean_width_dist_varyAlpha}).

The middle panels of Fig. \ref{fig:sc_mean_width_dist_varyAlpha} show the maximum Stokes numbers. The maximum Stokes number of the models with bouncing decrease with increasing turbulence for \(D=3\) and \(D=2.5\), although the decrease is only small for \(D=2.5\). For \(D=2\), the maximum Stokes number does not depend on \(\alpha\). Reaching the bouncing barrier already in the fractal growth regime for \(D=2\), is solely an effect of the decrease of bouncing velocity with mass, as the Stokes number and therefore the relative velocities stay constant with mass. Increasing the turbulence parameter does not have an effect on the Stokes number but on the relative velocities, which are increased. Therefore the bouncing velocity does not have to decrease as much as for low turbulence, and is therefore exceeded already at smaller masses. However, since the Stokes number is constant in mass, this does not change the Stokes number, at which the relative velocities are exceeded.

For the shape of the mass distribution (right panels in Fig. \ref{fig:sc_mean_width_dist_varyAlpha}), the behavior for different \(\alpha\) depends on the fractal dimensions. First, we look at the models with bouncing. For the non-fractal case \(D\!=\!3\), the mass distribution of the high turbulence model extends down to the lower bound of the mass domain as for the models without bouncing. The reason is that the fragmentation barrier is close to the bouncing barrier, such that some collisions already result in fragmentation (see Fig. \ref{fig:sc_sigma_d_highturb} in Appendix \ref{sec:appendix_plots}). For \(D \!=\! 2.5\), the same is true for the outer parts of the disk but in the inner parts, growth is stopped by the bouncing barrier already well below the fragmentation barrier. For \(D \!=\! 2\), this is true for the whole disk. Thus, when growth is bouncing limited, it is more difficult to still produce small particles for fractal models than for non-fractal models. The models without bouncing also show a peculiar behavior. For \(D \!=\!2.5\) and \(D\!=\!3\), the slope of \( \sigma_\mathrm{d}(m)\) becomes flatter when \(\alpha\) is decreased. For \(D=3\), the slope is even slightly negative. For \(D\!=\!2\), there is no such effect. However, in all cases the slopes are very flat.

\section{Advanced cross sections}
\label{sec:ac_models}
In Sect. \ref{sec:sc_models}, we applied a simple expression for the aerodynamical cross sections. Since this is a major simplification, we investigated how the results change for the different fractal dimensions if more elaborate expressions are applied. We refer to these models as advanced cross section (AC) models. In Sect. \ref{sec:ac_porosity_model}, we describe the advanced porosity model. Then, we compare the results of the AC models to that of the SC models for different fractal dimensions in Sect. \ref{sec:ac_vs_sc_models}.

\subsection{Porosity model}
\label{sec:ac_porosity_model}
\citet{Tazaki2021} derive semi-analytical expressions for the aerodynamical cross sections of fractal particles:
\begin{equation}
	\label{eq:Tazaki_A}
	A_\mathrm{T21} = N\pi a_\bullet^2 \begin{cases}
		\displaystyle 12.5 N^{-0.315} \exp\left(-\frac{2.53}{N^{0.0920}}\right)\ &\text{for } 1 \leq N < N_\mathrm{th},\\
		\displaystyle \frac{\chi}{1+(N-1)\tilde{\sigma}}\ &\text{for } N_\mathrm{th}\leq N,
	\end{cases}
\end{equation}
where \(N_\mathrm{th} = \min(11D - 8.5, 8.0)\) is a threshold, below which the cross section does not depend on the aggregation process and therefore not on the fractal dimension. \(\chi\) is a normalization factor connecting the \(N<N_\mathrm{th}\) and the \(N\geq N_\mathrm{th}\) regime, and \(\tilde{\sigma}\) is the overlapping efficiency, given by \citep{Tazaki2021}
\begin{equation}
	\label{eq:sigma_tilde}
	\tilde{\sigma} = \frac{\eta^{2/D}}{16}\int_\eta^\infty x^{-2/D}\exp(-x)\mathrm{d}x,
\end{equation}
where \(\eta = 2^{D-1}k_0/N\) and \(k_0=0.761(1-D)+\sqrt{3}\).

Unfortunately, the simple expression of Eq. \eqref{eq:simple_A} can not simply be replaced by Eq. \eqref{eq:Tazaki_A} while leaving everything else unchanged. The reason is the following. Using the aerodynamical cross sections given by Eq. \eqref{eq:Tazaki_A} together with the definition of radius as described by Eq. \eqref{eq:sc_radius} and the expression for the collision cross sections, given by Eq. \eqref{eq:A_coll}, would imply that the aerodynamical cross sections of one particle can be larger than its collision cross section with a small collision partner. This is unphysical, because the aerodynamical cross section corresponds to the collision cross section with a negligibly small collision partner (a gas molecule), and therefore always has to be smaller than the collision cross section with a collision partner of finite size.

The reason for this discrepancy is that Eq. \eqref{eq:Tazaki_A} implies that
initially all particles grow with a rather low actual fractal dimension \(D^\prime\), and approach the given fractal dimension \(D\) only at larger sizes. For example, even a \(D = 3\) aggregate gains some porosity initially (for \(N \lesssim 10^3\)) and only then proceeds growing with constant porosity.

Thus, to keep the model consistent when using the aerodynamical cross sections given by Eq. \eqref{eq:Tazaki_A}, we have to redefine the radius of a fractal particle. We chose to take the area-equivalent radius. The full expression for the radius as a function of mass is then given by
\begin{align}
	\label{eq:ac_radius}
	a(m) = \begin{cases}
		\displaystyle a_\bullet \left(\frac{m}{m_\bullet}\right)^{1/3}\  & \text{for } m < m_\bullet,\\
		\displaystyle \sqrt{A_\mathrm{T21}(m)/\pi} & \text{for } m_\bullet \leq m \leq m_\mathrm{mff},\\
		\displaystyle a_\bullet \phi_\mathrm{min}^{-1/3}\left(\frac{m}{m_\bullet}\right)^{1/3}\  & \text{for }  m_\mathrm{mff} < m,
		\end{cases}
\end{align}
where \(m_\mathrm{mff}\) is the mass, at which the minimum filling factor is reached.

The advantage of this choice is that the complications are all absorbed into the definition of the radius. Consequently, the aerodynamical cross section can still be expressed as \(A_\mathrm{aero}=\pi a^2\) and  the expression for the Stokes number (see Eq. \eqref{eq:Stokes}) also stays the same. The collision cross section might be underestimated for large collision partners, but for small collision partners, it approaches the aerodynamical cross section, which is what one would expect.

\begin{figure}
   \centering
   \includegraphics[width=\hsize]{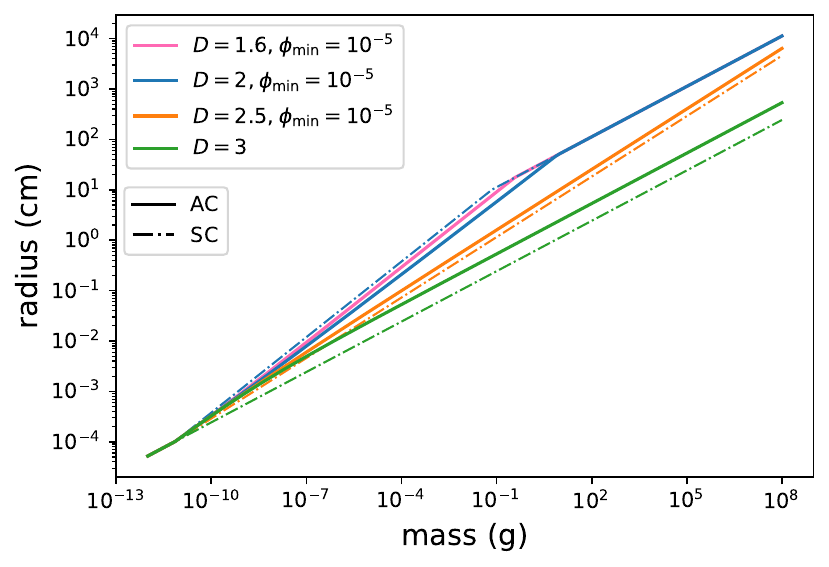}
      \caption{Mass-radius relations of the AC models. For comparison, also the SC models are plotted as dash-dotted lines.}
         \label{fig:ac_m_a_relations}
\end{figure}

The filling factor is then given by
\begin{align}
	\label{eq:ac_phi}
	\phi = \begin{cases}
		\displaystyle 1 \quad & \text{for } m < m_\bullet,\\
		\displaystyle \max\left[\frac{3m\sqrt{\pi}}{4 A_\mathrm{T21}(m)^{3/2}\rho_\mathrm{s}}, \phi_\mathrm{min}\right]& \text{else}.
	\end{cases}
\end{align}
We refer to models using the porosity prescription described above as the advanced cross section (AC) models. The mass-radius relations of the AC models are shown in Fig. \ref{fig:ac_m_a_relations}.
\subsection{Comparison between AC and SC models}
\label{sec:ac_vs_sc_models}

\begin{table}
	\centering
	\caption{Properties of the AC models}
	\label{tab:ac_models}
	\begin{tabular}{lllll}
\hline\hline
Model name  & \(D\) &\(\phi_\mathrm{min}\) &\(\alpha\)	& bo.? \\

\hline

\model{AC\_D3\_a4}     				& 3		& -	 			& \(10^{-4}\) & yes \\
\model{AC\_D25\_fm5\_a4} 			& 2.5	& \(10^{-5}\)	& \(10^{-4}\) & yes \\
\model{AC\_D2\_fm5\_a4}				& 2		& \(10^{-5}\)	& \(10^{-4}\) & yes \\
\model{AC\_D16\_fm5\_a4}				& 1.6	& \(10^{-5}\)	& \(10^{-4}\) & yes \\

\hline

\model{AC\_D3\_a4\_nobo}				& 3		& -		 		& \(10^{-4}\) & no \\
\model{AC\_D25\_fm5\_a4\_nobo}		& 2.5	& \(10^{-5}\)	& \(10^{-4}\) & no \\
\model{AC\_D2\_fm5\_a4\_nobo}		& 2		& \(10^{-5}\)	& \(10^{-4}\) & no \\
\model{AC\_D16\_fm5\_a4\_nobo}		& 1.6	& \(10^{-5}\)	& \(10^{-4}\) & no \\
\hline
\end{tabular}
	\tablefoot{Same as Table \ref{tab:sc_models} but for the AC models. The model names are constructed in the same way, except that they start with \model{AC} instead of \model{SC}.}
\end{table}

\begin{figure}
   \centering
   \includegraphics[width=\hsize]{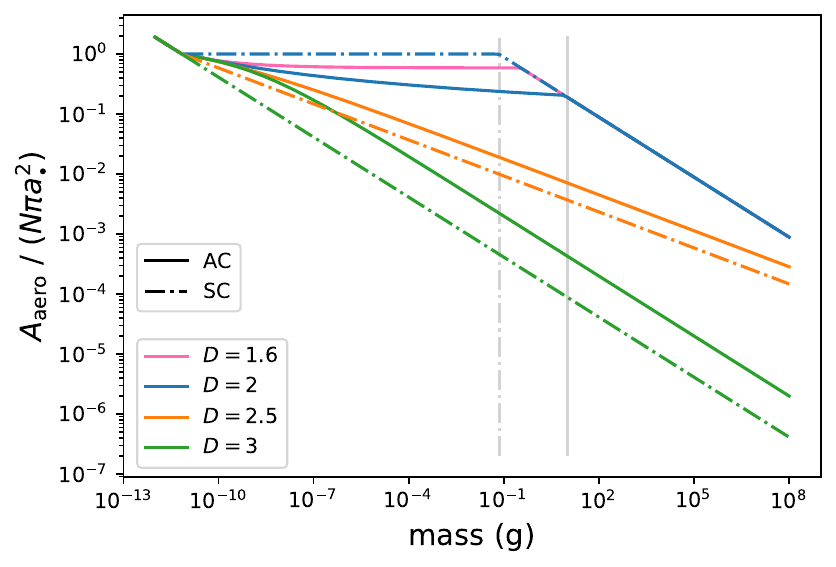}
      \caption{Comparison between aerodynamical cross sections of the AC models (solid lines) and the SC models (dash-dotted lines). The cross sections are normalized to the sum of the monomer cross sections, \(N\pi a_\bullet^2\). The vertical lines indicate the transition to non-fractal growth for the \(D=2\) models.}
         \label{fig:ac_vs_sc_cross_sections}
\end{figure}

\begin{figure*}
   \centering
   \includegraphics[width=\hsize]{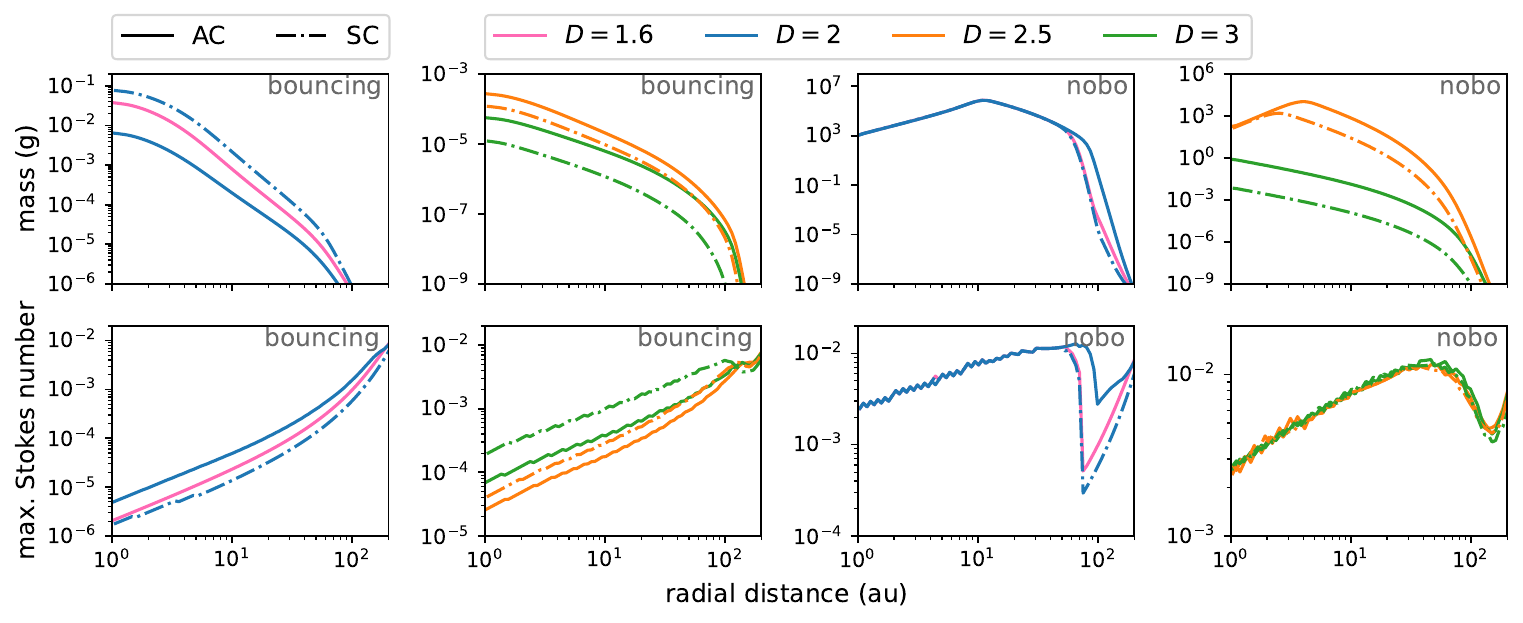}
      \caption{Comparison between AC models (solid lines) and SC models (dash-dotted lines) for different fractal dimensions. Upper row: Mean mass. Lower row: Maximum Stokes number. Left: Models with bouncing. Right: Models without bouncing. For better visibility, we have drawn the models with \(D\leq2\) and \(D>2\) into different panels.}
         \label{fig:ac_mean}
\end{figure*}

To investigate the effect of using the advanced cross section expressions, we ran eight AC models with \(\phi_\mathrm{min}=10^{-5}\), \(\alpha=10^{-4}\), and different fractal dimensions both with and without bouncing. The models and their properties are listed in Table \ref{tab:ac_models}.

Figure \ref{fig:ac_vs_sc_cross_sections} compares the advanced cross sections to the simple cross sections. For \(D=3\) and \(D=2.5\) the advanced cross sections are larger than the simple cross sections. This is because the actual fractal dimension in the AC models is initially smaller than the given fractal dimension (see Sect. \ref{sec:ac_porosity_model}). For \(D=2\), the ratio \(A_\mathrm{aero} / (N\pi a_\bullet^2) \!=\! 1\) for the simple cross sections. Hence, the cross section of the aggregate is exactly equal to the sum of the cross sections of all monomers that the aggregate consists of, which corresponds to the maximum possible cross section. Thus, there is no overlapping of monomer cross sections, such that one monomer would insulate parts of another monomer from the gas flow. Instead, all monomers fully face the gas flow. 

In the case of the advanced cross section, such overlapping effects are considered \citep[for details, see][]{Tazaki2021}. The total cross section of the aggregate is therefore somewhat smaller than the sum of the cross sections of all monomers. For this reason, also values \(D \!<\!  2\) are possible. Decreasing \(D\) below 2 makes the overlapping less efficient and the cross sections get closer to the maximum possible cross section \(N\pi a_\bullet^2\), which corresponds to the simple cross section with \(D=2\).

Fig. \ref{fig:ac_mean} shows the comparison of maximum Stokes number and the mean mass between AC and SC models. 

In the fragmentation-limited case, the maximum Stokes numbers reached by the particles are independent of the cross section model, because the fragmentation barrier is only determined by the Stokes number. For \(D=2\), also the mean mass is independent of the cross section model (except in the very outer part of the disk), because the fragmentation barrier lies in the non-fractal growth regime for \(D=2\), where the advanced and simple cross sections do not differ. For \(D=2.5\), the difference in mean mass is very small in the inner parts of the disk, where the particles are in the Stokes regime. In the outer parts of the disk, the mean mass of the AC model is \mbox{\(\sim\!20\times\)} larger. For \(D\!=\!3\), the difference is about a factor of \mbox{\(\sim\!100\)}. In both cases, the difference in mean mass is an effect of the difference in mass dependence of the Stokes numbers between AC and SC models. In the Epstein regime, \(\mathrm{St}\propto N/A_\mathrm{aero}\), therefore, the mass differences between AC and SC models for a given \(A_\mathrm{aero}/N\), which can be read from Fig. \ref{fig:ac_vs_sc_cross_sections}, correspond to the mass differences of the fragmentation-limited distribution. 

In the bouncing-limited case, the behavior of mean mass and maximum Stokes number depends on the fractal dimension. In the case of \(D=2\), for a fixed mass, the advanced cross sections are smaller than the simple cross sections. Thus, despite the same value of \(D\), the AC models can be viewed as having a "lower fractality" than the SC models. As it was already shown in Sects. \ref{sec:sc_varyD} and \ref{sec:sc_varyFillmin}, this leads to smaller masses for the AC models but larger maximum Stokes numbers due to the mass dependence of the bouncing velocity. For \(D=2.5\) and \(D=3\), the AC models have a "higher fractality" than the SC models, and therefore reach larger masses but smaller maximum Stokes numbers.

In all cases, decreasing the fractal dimension of the AC models from \(D=2\) to \(D=1.6\), decreases the difference to the SC model with \(D=2\).

\section{Discussion}
\label{sec:discussion}
In the following, we discuss some aspects of our models and the results. First, we discuss how fractal growth changes the Stokes numbers that are reached at the end of the simulation, and what the implications are in the context of the streaming instability. Then, we discuss our finding that growth becomes slower with smaller fractal dimensions and try to explain differences to previous studies. 

\subsection{Bouncing and fragmentation velocities}
\label{sec:collision_discussion}
Due to the large uncertainties in the collision behavior of porous particles and to keep things simple, we assumed that the collision outcome is not affected by the filling factor. Our expression for the bouncing velocity (Eqs. \eqref{eq:v_s} and \eqref{eq:v_b}) is based on the model of \citet{Guettler2010}, where the bouncing velocity decreases with mass. This behavior is also found in more recent laboratory \citep[e.g.,][]{Schraepler2022} and numerical \citep[e.g.,][]{Oshiro2025} studies, although the exact mass dependence of \(v_\mathrm{b}\) is somewhat different.

Our assumption that the collision outcome does not depend on filling factor is a major simplification. \citet{Oshiro2025} investigate the bouncing behavior of porous aggregates numerically and find that the bouncing velocity increases with decreasing filling factor. Other numerical studies suggest that bouncing does not occur if the filling factor of a particle is below \(\sim\!\!0.3 - 0.4\) \citep{Wada2011,SeizingerKley2013}, while laboratory studies still find bouncing for lower filling factors \citep[e.g.,][]{BlumMuench1993,Langkowski2008,Schraepler2022}. This discrepancy might be solved by a size dependency of the threshold filling factor \citep{Arakawa2023}. It should be noted that the filling factors explored in these laboratory experiments are still much larger than the filling factors reached in this study.

An increasing bouncing threshold velocity for decreasing filling factors would lead to even larger masses for fractal particles in bouncing-limited mass distributions and the effect of smaller maximum Stokes numbers for smaller fractal dimensions would be reduced or even reversed, depending of the exact dependence of the bouncing velocity on the filling factor. A critical filling factor below which bouncing does not occur would either lead to a fragmentation-limited mass distribution if the critical filling factor is reached before the bouncing barrier, or to a bouncing-limited mass distribution if the bouncing barrier is reached before the critical filling factor.

The fragmentation velocity might also depend on the fractal dimension or filling factor. \citet{Wada2009} find that the fragmentation velocity decreases by a factor ~2 when the fractal dimension is decreased from \(D=3\) to \(D=2\). \citet{Gunkelmann2016} find that for a fixed mass the fragmentation velocity decreases with decreasing filling factor for \(\phi\lesssim0.1\). It should be noted that the threshold velocities above which fragmentation occurs in these studies are on the order of a few 10 m/s, i.e., much larger than the fragmentation velocity assumed in this study. 

A decreasing fragmentation velocity with decreasing fractal dimension would reduce the effect that smaller fractal dimensions reach larger masses and would lead to a decrease of maximum Stokes number for small fractal dimensions.

\subsection{Streaming instability}
We showed that with our choice of the fragmentation velocity \(v_\mathrm{f} \!=\! 100\ \mathrm{cm\ s^{-1}}\) particles can not reach planetesimal masses even if their growth is highly fractal. Therefore, other mechanisms such as the streaming instability \citep{YoudinGoodman2005,JohansenYoudin2007,Johansen2007} must be invoked to explain the formation of planetesimals. Studies have found that the Stokes numbers must exceed \mbox{\(\mathrm{St} \!\sim\! 10^{-3}\)} for the streaming instability to occur \citep{Carrera2015,Yang2017,LiYoudin2021}. The exact value depends on the dust-to-gas ratio \(\Sigma_\mathrm{d}/\Sigma_\mathrm{g}\), which generally has to be \(\gtrsim\!0.005\) \citep{LiYoudin2021} or \(\gtrsim\!0.01\) \citep{Carrera2015,Yang2017}.

\begin{figure}
   \centering
   \includegraphics[width=\hsize]{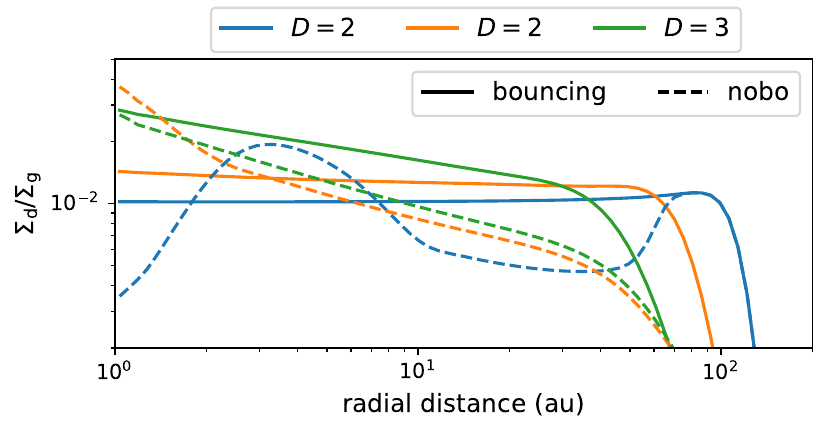}
      \caption{Dust-to-gas ratio of the SC models with bouncing (solid) and without bouncing (dashed) for different fractal dimensions.}
         \label{fig:St_varyD}
\end{figure}

For the models with bouncing, the dust-to-gas ratio is somewhat larger for larger fractal dimensions, because radial drift is more efficient and dust drifting inward from the outer regions increases the dust-to-gas ratio in the inner regions. For the models without bouncing, the picture is less distinct. However, the dust-to-gas ratios are similar for all fractal dimensions except in the inner regions (\(r\lesssim3\) au), where the dust-to-gas ratio of the model with \(D=2\) is significantly lower (see Fig. \ref{fig:St_varyD}). Note that the dust-to-gas ratio can be enhanced by dust traps \citep[e.g.,][]{KretkeLin2007,Dullemond2018}, which we do not include in this study.

As shown in Sect. \ref{sec:sc_varyD}, the maximum Stokes number that is reached does not depend on the fractal dimension in the fragmentation-limited case. For all fractal dimensions, the maximum Stokes numbers are above \(10^{-3}\). In the bouncing-limited case, a lower fractal dimension even leads to smaller Stokes numbers. While maximum Stokes numbers of the models with \(D=3\) and \(D=2.5\) reach the threshold of \(10^{-3}\) in the outer parts of the disk, the maximum Stokes numbers of the model with \(D=2\) is much smaller than \(10^{-3}\) in the whole disk. Thus, under the assumption that the effect of fractal growth on the bouncing and fragmentation threshold velocities is negligible, fractal growth is not beneficial for the streaming instability in the fragmentation-limited case, and even disadvantageous in the bouncing-limited case. However, as discussed in Sect. \ref{sec:collision_discussion}, the bouncing velocity might be larger for fractal particles, or bouncing might not occur below a critical filling factor. In this case, fractal growth could be less disadvantageous or even beneficial for the streaming instability. If the fragmentation threshold velocity was somewhat smaller for fractal particles, fractal growth would be disadvantageous for the streaming instability in the fragmentation-limited case.

\subsection{Fractal growth rate}
\label{sec:discussion_time_evolution}

\begin{figure}
    \centering
    \includegraphics[width=\hsize]{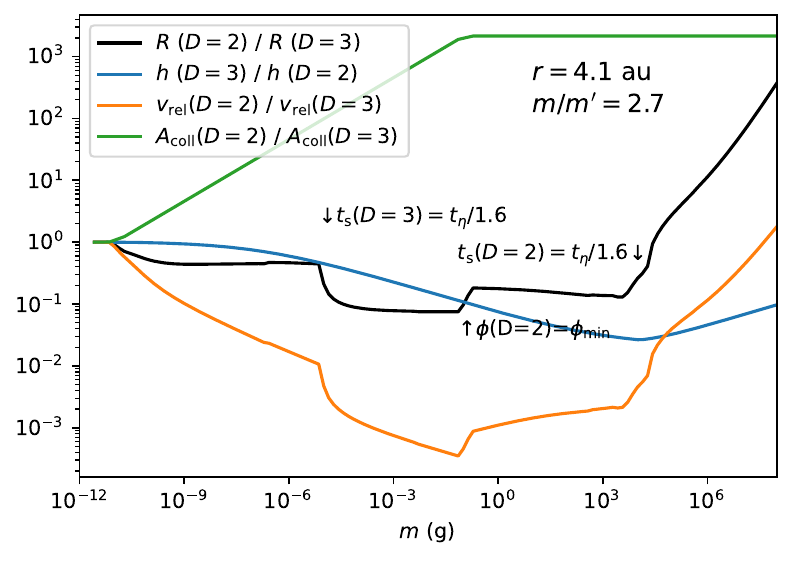}
    \caption{Contributions of the settling factor (blue), the relative velocitiy (orange) and the collision cross section (green) to the ratio of the growth rates \(R\) (black) between the SC models with \(D=2\) and \(D=3\) for a mass ratio of 2.7.}
    \label{fig:growth_rate_contributions}
\end{figure}

We found that starting with the same initial mass distribution, particles in a model with a large fractal dimension will reach larger masses in a certain amount of time than the particles in a model with a small fractal dimension. This is in contrast to results of previous studies that found that fractal or porous particles grow faster \citep{DullemondDominik2005,Okuzumi2012,Estrada2022,Michoulier2024}. Although \citet{Okuzumi2012} focus on the growth time scales in the Stokes regime, the fractal particles in their models grow somewhat faster than the non-fractal counterparts already in the Epstein regime (see their Fig. 4 and Fig. 7). In the following, we discuss the different results in our and other studies. 

When comparing the growth rate for two different fractal dimensions, several counteracting effects play a role. For a smaller fractal dimension, the collision cross section becomes larger, which increases the growth rate, but at the same time the relative velocities become smaller (as long as \(\mathrm{St}<1\)) and the dust scale height becomes larger, which decreases the growth rate. 

Figure \ref{fig:growth_rate_contributions} shows how the different effects contribute to the total ratio between the growth rates of the models with \(D=3\) and \(D=2\). For small masses, the combined effect of relative velocity and settling is slightly stronger than the cross section effect, consistent with Eq. \eqref{eq:R}. However, this might be very sensitive to the model specifics, since the cross section effect and the relative velocity effect almost cancel out. Somewhat different treatments of settling, relative velocities or collision cross sections than in our model might let the cross section effect become slightly larger than the relative velocity and settling effect, such that fractal particles would grow faster than non-fractal particles. When the non-fractal model leaves the \(t_s < t_\eta\) regime (due to a transition regime, this already happens at \(t_\eta/1.6\)), the relative velocity increases, which intensifies the relative velocity difference between \(D=2\) and \(D=3\). When the \(D=2\) model reaches the non-fractal growth regime, the difference in relative velocities is decreased again. When the \(D=2\) model leaves the \(t_\mathrm{s} < t_\eta\) regime, its relative velocities rapidly increase and consequently, the growth rate becomes larger than that of the \(D=3\) model. However, the \(D=3\) model reaches the growth barrier already at much smaller masses, such that the comparison of growth rate at these mass regimes is not meaningful in regard to our models.

\section{Conclusions}
\label{sec:conclusions}
In this work, we have modeled dust growth including a prescription for porosity of fractal dust particles, and investigated the dust evolution for different degrees of fractality, i.e., for different fractal dimensions. We have run models with and without bouncing and we have applied two kinds of cross sections: the cross section of a spherical particle, and a more advanced, semi-analytic expression. This lead us to the following conclusions:

\begin{itemize}

	\item Decreasing the fractal dimension shifts the fragmentation and bouncing barriers towards larger masses. Therefore, models with smaller fractal dimensions reach larger masses, with and without bouncing.
	
	\item For a fragmentation-limited mass distribution, the maximum Stokes numbers that are reached do not depend on the fractal dimension. The mean Stokes numbers show a small dependence on fractal dimension. Therefore, under the assumption that the fragmentation velocity does not depend on the fractal dimension or filling factor, fractal growth is not beneficial for the streaming instability in the case of a fragmentation-limited mass distribution.
	
	\item For a bouncing-limited mass distribution, both the maximum and mean Stokes numbers decrease with decreasing fractal dimension. Therefore, under the assumption that the bouncing velocity does not depend on the fractal dimension or filling factor, fractal growth is disadvantageous for the streaming instability in the case of a bouncing-limited mass distribution.
	
	\item Fractal growth is slower than non-fractal growth, because the larger collision cross sections are compensated by the smaller relative velocities and degree of vertical settling. However the effect is only small. Changing the model specifics might lead to different results.
	
	\item In the high turbulence case (\(\alpha=10^{-3}\)), the bouncing and fragmentation barrier are close together for non-fractal particles, such that some fragmentation events occur, although bouncing is active. For moderately (\(D=2.5\)) or highly fractal (\(D=2\)) particles, this only occurs in the outer parts of the disk or not at all, respectively, because the bouncing and fragmentation barriers are further apart.

	\item Using the more advanced expressions for the cross sections does not change the maximum Stokes number that the particles reach in the fragmentation-limited case, while it leads to larger maximum Stokes numbers for \(D=2\) and to smaller maximum Stokes numbers for \(D=2.5\) and \(D=3\). The masses that are reached with the advanced cross sections become smaller for \(D=2\) and larger for \(D=2.5\) and \(D=3\). 
	
	\item The \(D=2\) case of the simple cross sections corresponds to the maximum possible cross sections. Highly fractal (\(D<2\)) models with advanced cross sections therefore approach the \(D=2\) model with simple cross sections, as the fractal dimension is decreased.
	
\end{itemize}

\begin{acknowledgements} 
	We acknowledge support by the High Performance and Cloud Computing Group at the Zentrum für
	Datenverarbeitung of the University of Tübingen, the state of Baden-Württemberg through bwHPC and the
	German Research Foundation (DFG) through grant INST 37/935-1 FUGG.
    Carsten Dominik acknowledges support from the ERC AdG project grant 101199769 GT4Pebbles.
\end{acknowledgements}

\bibliographystyle{aa} 
\bibliography{References.bib} 

@ARTICLE{Arakawa2023,
       author = {{Arakawa}, Sota and {Okuzumi}, Satoshi and {Tatsuuma}, Misako and {Tanaka}, Hidekazu and {Kokubo}, Eiichiro and {Nishiura}, Daisuke and {Furuichi}, Mikito and {Nakamoto}, Taishi},
        title = "{Size Dependence of the Bouncing Barrier in Protoplanetary Dust Growth}",
      journal = {\apjl},
         year = 2023,
        month = jul,
       volume = {951},
       number = {1},
          eid = {L16},
        pages = {L16},
          doi = {10.3847/2041-8213/acdb5f},
archivePrefix = {arXiv},
       eprint = {2306.04070},
 primaryClass = {astro-ph.EP},
       adsurl = {https://ui.adsabs.harvard.edu/abs/2023ApJ...951L..16A}
}

@ARTICLE{BeutelDullemond2023,
author = {{Beutel}, M. and {Dullemond}, C.~P.},
title = "{An improved Representative Particle Monte Carlo method for the simulation of particle growth}",
journal = {\aap},
year = 2023,
month = feb,
volume = {670},
eid = {A134},
pages = {A134},
doi = {10.1051/0004-6361/202244955},
adsurl = {ui.adsabs.harvard.edu/abs/2023A&A...670A.134B}
}

@ARTICLE{Birnstiel2009,
       author = {{Birnstiel}, T. and {Dullemond}, C.~P. and {Brauer}, F.},
        title = "{Dust retention in protoplanetary disks}",
      journal = {\aap},
         year = 2009,
        month = aug,
       volume = {503},
       number = {1},
        pages = {L5-L8},
          doi = {10.1051/0004-6361/200912452},
archivePrefix = {arXiv},
       eprint = {0907.0985},
 primaryClass = {astro-ph.EP},
       adsurl = {https://ui.adsabs.harvard.edu/abs/2009A&A...503L...5B}
}

@ARTICLE{Birnstiel2012,
       author = {{Birnstiel}, T. and {Klahr}, H. and {Ercolano}, B.},
        title = "{A simple model for the evolution of the dust population in protoplanetary disks}",
      journal = {\aap},
         year = 2012,
        month = mar,
       volume = {539},
          eid = {A148},
        pages = {A148},
          doi = {10.1051/0004-6361/201118136},
archivePrefix = {arXiv},
       eprint = {1201.5781},
 primaryClass = {astro-ph.EP},
       adsurl = {https://ui.adsabs.harvard.edu/abs/2012A&A...539A.148B}
}

@ARTICLE{BlumMuench1993,
       author = {{Blum}, J{\"u}rgen and {M{\"u}nch}, Michael},
        title = "{Experimental Investigations on Aggregate-Aggregate Collisions in the Early Solar Nebula}",
      journal = {\icarus},
         year = 1993,
        month = nov,
       volume = {106},
       number = {1},
        pages = {151-167},
          doi = {10.1006/icar.1993.1163},
       adsurl = {https://ui.adsabs.harvard.edu/abs/1993Icar..106..151B}
}

@ARTICLE{BlumWurm2008,
author = {{Blum}, J. and {Wurm}, G.},
title = "{The growth mechanisms of macroscopic bodies in protoplanetary disks.}",
journal = {\araa},
year = 2008,
month = sep,
volume = {46},
pages = {21-56},
doi = {10.1146/annurev.astro.46.060407.145152},
adsurl = {ui.adsabs.harvard.edu/abs/2008ARA&A..46...21B}
}

@ARTICLE{Brauer2008,
       author = {{Brauer}, F. and {Dullemond}, C.~P. and {Henning}, Th.},
        title = "{Coagulation, fragmentation and radial motion of solid particles in protoplanetary disks}",
      journal = {\aap},
         year = 2008,
        month = mar,
       volume = {480},
       number = {3},
        pages = {859-877},
          doi = {10.1051/0004-6361:20077759},
archivePrefix = {arXiv},
       eprint = {0711.2192},
 primaryClass = {astro-ph},
       adsurl = {https://ui.adsabs.harvard.edu/abs/2008A&A...480..859B}
}

@ARTICLE{Carrera2015,
       author = {{Carrera}, Daniel and {Johansen}, Anders and {Davies}, Melvyn B.},
        title = "{How to form planetesimals from mm-sized chondrules and chondrule aggregates}",
      journal = {\aap},
         year = 2015,
        month = jul,
       volume = {579},
          eid = {A43},
        pages = {A43},
          doi = {10.1051/0004-6361/201425120},
archivePrefix = {arXiv},
       eprint = {1501.05314},
 primaryClass = {astro-ph.EP},
       adsurl = {https://ui.adsabs.harvard.edu/abs/2015A&A...579A..43C}
}

@ARTICLE{DominikDullemond2024,
       author = {{Dominik}, C. and {Dullemond}, C.~P.},
        title = "{The bouncing barrier revisited: Impact on key planet formation processes and observational signatures}",
      journal = {\aap},
         year = 2024,
        month = feb,
       volume = {682},
          eid = {A144},
        pages = {A144},
          doi = {10.1051/0004-6361/202347716},
archivePrefix = {arXiv},
       eprint = {2312.06000},
 primaryClass = {astro-ph.EP},
       adsurl = {https://ui.adsabs.harvard.edu/abs/2024A&A...682A.144D}
}

@ARTICLE{DominikTielens1997,
       author = {{Dominik}, C. and {Tielens}, A.~G.~G.~M.},
        title = "{The Physics of Dust Coagulation and the Structure of Dust Aggregates in Space}",
      journal = {\apj},
         year = 1997,
        month = may,
       volume = {480},
       number = {2},
        pages = {647-673},
          doi = {10.1086/303996},
       adsurl = {https://ui.adsabs.harvard.edu/abs/1997ApJ...480..647D}
}

@ARTICLE{Dubrulle1995,
       author = {{Dubrulle}, B. and {Morfill}, G. and {Sterzik}, M.},
        title = "{The dust subdisk in the protoplanetary nebula.}",
      journal = {\icarus},
         year = 1995,
        month = apr,
       volume = {114},
       number = {2},
        pages = {237-246},
          doi = {10.1006/icar.1995.1058},
       adsurl = {https://ui.adsabs.harvard.edu/abs/1995Icar..114..237D}
}

@ARTICLE{DullemondDominik2005,
       author = {{Dullemond}, C.~P. and {Dominik}, C.},
        title = "{Dust coagulation in protoplanetary disks: A rapid depletion of small grains}",
      journal = {\aap},
         year = 2005,
        month = may,
       volume = {434},
       number = {3},
        pages = {971-986},
          doi = {10.1051/0004-6361:20042080},
archivePrefix = {arXiv},
       eprint = {astro-ph/0412117},
 primaryClass = {astro-ph},
       adsurl = {https://ui.adsabs.harvard.edu/abs/2005A&A...434..971D}
}

@ARTICLE{Dullemond2018,
       author = {{Dullemond}, Cornelis P. and {Birnstiel}, Tilman and {Huang}, Jane and {Kurtovic}, Nicol{\'a}s T. and {Andrews}, Sean M. and {Guzm{\'a}n}, Viviana V. and {P{\'e}rez}, Laura M. and {Isella}, Andrea and {Zhu}, Zhaohuan and {Benisty}, Myriam and et al.},
        title = "{The Disk Substructures at High Angular Resolution Project (DSHARP). VI. Dust Trapping in Thin-ringed Protoplanetary Disks}",
      journal = {\apjl},
         year = 2018,
        month = dec,
       volume = {869},
       number = {2},
          eid = {L46},
        pages = {L46},
          doi = {10.3847/2041-8213/aaf742},
archivePrefix = {arXiv},
       eprint = {1812.04044},
 primaryClass = {astro-ph.EP},
       adsurl = {https://ui.adsabs.harvard.edu/abs/2018ApJ...869L..46D}
}

@ARTICLE{EstradaCuzzi2008,
author = {{Estrada}, P.~R. and {Cuzzi}, J.~N.},
title = "{Solving the Coagulation Equation by the Moments Method}",
journal = {\apj},
year = 2008,
month = jul,
volume = {682},
number = {1},
pages = {515-526},
doi = {10.1086/589685},
archivePrefix = {arXiv},
eprint = {0804.4189},
primaryClass = {astro-ph},
adsurl = {ui.adsabs.harvard.edu/abs/2008ApJ...682..515E}
}

@ARTICLE{Estrada2022,
       author = {{Estrada}, Paul R. and {Cuzzi}, Jeffrey N. and {Umurhan}, Orkan M.},
        title = "{Global Modeling of Nebulae with Particle Growth, Drift, and Evaporation Fronts. II. The Influence of Porosity on Solids Evolution}",
      journal = {\apj},
         year = 2022,
        month = sep,
       volume = {936},
       number = {1},
          eid = {42},
        pages = {42},
          doi = {10.3847/1538-4357/ac7ffd},
archivePrefix = {arXiv},
       eprint = {2207.12626},
 primaryClass = {astro-ph.EP},
       adsurl = {https://ui.adsabs.harvard.edu/abs/2022ApJ...936...42E}
}

@ARTICLE{Guettler2010,
       author = {{G{\"u}ttler}, C. and {Blum}, J. and {Zsom}, A. and {Ormel}, C.~W. and {Dullemond}, C.~P.},
        title = "{The outcome of protoplanetary dust growth: pebbles, boulders, or planetesimals?. I. Mapping the zoo of laboratory collision experiments}",
      journal = {\aap},
         year = 2010,
        month = apr,
       volume = {513},
          eid = {A56},
        pages = {A56},
          doi = {10.1051/0004-6361/200912852},
archivePrefix = {arXiv},
       eprint = {0910.4251},
 primaryClass = {astro-ph.EP},
       adsurl = {https://ui.adsabs.harvard.edu/abs/2010A&A...513A..56G}
}

@ARTICLE{Guettler2019,
       author = {{G{\"u}ttler}, C. and {Mannel}, T. and {Rotundi}, A. and {Merouane}, S. and {Fulle}, M. and {Bockel{\'e}e-Morvan}, D. and {Lasue}, J. and {Levasseur-Regourd}, A.~C. and {Blum}, J. and {Naletto}, G. and {Sierks}, H. and {Hilchenbach}, M. and {Tubiana}, C. and {Capaccioni}, F. and {Paquette}, J.~A. and {Flandes}, A. and {Moreno}, F. and {Agarwal}, J. and {Bodewits}, D. and {Bertini}, I. and {Tozzi}, G.~P. and {Hornung}, K. and {Langevin}, Y. and {Kr{\"u}ger}, H. and {Longobardo}, A. and {Della Corte}, V. and {T{\'o}th}, I. and {Filacchione}, G. and {Ivanovski}, S.~L. and {Mottola}, S. and {Rinaldi}, G.},
        title = "{Synthesis of the morphological description of cometary dust at comet 67P/Churyumov-Gerasimenko}",
      journal = {\aap},
         year = 2019,
        month = oct,
       volume = {630},
          eid = {A24},
        pages = {A24},
          doi = {10.1051/0004-6361/201834751},
archivePrefix = {arXiv},
       eprint = {1902.10634},
 primaryClass = {astro-ph.EP},
       adsurl = {https://ui.adsabs.harvard.edu/abs/2019A&A...630A..24G}
}

@ARTICLE{Gunkelmann2016,
       author = {{Gunkelmann}, Nina and {Ringl}, Christian and {Urbassek}, Herbert M.},
        title = "{Influence of porosity on collisions between dust aggregates}",
      journal = {\aap},
         year = 2016,
        month = may,
       volume = {589},
          eid = {A30},
        pages = {A30},
          doi = {10.1051/0004-6361/201628081},
       adsurl = {https://ui.adsabs.harvard.edu/abs/2016A&A...589A..30G}
}

@ARTICLE{Johansen2007,
       author = {{Johansen}, Anders and {Oishi}, Jeffrey S. and {Mac Low}, Mordecai-Mark and {Klahr}, Hubert and {Henning}, Thomas and {Youdin}, Andrew},
        title = "{Rapid planetesimal formation in turbulent circumstellar disks}",
      journal = {\nat},
         year = 2007,
        month = aug,
       volume = {448},
       number = {7157},
        pages = {1022-1025},
          doi = {10.1038/nature06086},
archivePrefix = {arXiv},
       eprint = {0708.3890},
 primaryClass = {astro-ph},
       adsurl = {https://ui.adsabs.harvard.edu/abs/2007Natur.448.1022J}
}

@ARTICLE{JohansenYoudin2007,
       author = {{Johansen}, A. and {Youdin}, A.},
        title = "{Protoplanetary Disk Turbulence Driven by the Streaming Instability: Nonlinear Saturation and Particle Concentration}",
      journal = {\apj},
         year = 2007,
        month = jun,
       volume = {662},
       number = {1},
        pages = {627-641},
          doi = {10.1086/516730},
archivePrefix = {arXiv},
       eprint = {astro-ph/0702626},
 primaryClass = {astro-ph},
       adsurl = {https://ui.adsabs.harvard.edu/abs/2007ApJ...662..627J}
}

@ARTICLE{Kataoka2013a,
       author = {{Kataoka}, A. and {Tanaka}, H. and {Okuzumi}, S. and {Wada}, K.},
        title = "{Static compression of porous dust aggregates}",
      journal = {\aap},
         year = 2013,
        month = jun,
       volume = {554},
          eid = {A4},
        pages = {A4},
          doi = {10.1051/0004-6361/201321325},
archivePrefix = {arXiv},
       eprint = {1303.3351},
 primaryClass = {astro-ph.EP},
       adsurl = {https://ui.adsabs.harvard.edu/abs/2013A&A...554A...4K}
}

@ARTICLE{Kataoka2013b,
       author = {{Kataoka}, Akimasa and {Tanaka}, Hidekazu and {Okuzumi}, Satoshi and {Wada}, Koji},
        title = "{Fluffy dust forms icy planetesimals by static compression}",
      journal = {\aap},
         year = 2013,
        month = sep,
       volume = {557},
          eid = {L4},
        pages = {L4},
          doi = {10.1051/0004-6361/201322151},
archivePrefix = {arXiv},
       eprint = {1307.7984},
 primaryClass = {astro-ph.EP},
       adsurl = {https://ui.adsabs.harvard.edu/abs/2013A&A...557L...4K}
}

@ARTICLE{KretkeLin2007,
       author = {{Kretke}, Katherine A. and {Lin}, D.~N.~C.},
        title = "{Grain Retention and Formation of Planetesimals near the Snow Line in MRI-driven Turbulent Protoplanetary Disks}",
      journal = {\apjl},
         year = 2007,
        month = jul,
       volume = {664},
       number = {1},
        pages = {L55-L58},
          doi = {10.1086/520718},
archivePrefix = {arXiv},
       eprint = {0706.1272},
 primaryClass = {astro-ph},
       adsurl = {https://ui.adsabs.harvard.edu/abs/2007ApJ...664L..55K}
}

@ARTICLE{Krijt2015,
       author = {{Krijt}, S. and {Ormel}, C.~W. and {Dominik}, C. and {Tielens}, A.~G.~G.~M.},
        title = "{Erosion and the limits to planetesimal growth}",
      journal = {\aap},
         year = 2015,
        month = feb,
       volume = {574},
          eid = {A83},
        pages = {A83},
          doi = {10.1051/0004-6361/201425222},
archivePrefix = {arXiv},
       eprint = {1412.3593},
 primaryClass = {astro-ph.EP},
       adsurl = {https://ui.adsabs.harvard.edu/abs/2015A&A...574A..83K}
}

@ARTICLE{Langkowski2008,
       author = {{Langkowski}, Doreen and {Teiser}, Jens and {Blum}, J{\"u}rgen},
        title = "{The Physics of Protoplanetesimal Dust Agglomerates. II. Low-Velocity Collision Properties}",
      journal = {\apj},
         year = 2008,
        month = mar,
       volume = {675},
       number = {1},
        pages = {764-776},
          doi = {10.1086/525841},
archivePrefix = {arXiv},
       eprint = {0711.2148},
 primaryClass = {astro-ph},
       adsurl = {https://ui.adsabs.harvard.edu/abs/2008ApJ...675..764L}
}

@ARTICLE{LiYoudin2021,
       author = {{Li}, Rixin and {Youdin}, Andrew N.},
        title = "{Thresholds for Particle Clumping by the Streaming Instability}",
      journal = {\apj},
         year = 2021,
        month = oct,
       volume = {919},
       number = {2},
          eid = {107},
        pages = {107},
          doi = {10.3847/1538-4357/ac0e9f},
archivePrefix = {arXiv},
       eprint = {2105.06042},
 primaryClass = {astro-ph.EP},
       adsurl = {https://ui.adsabs.harvard.edu/abs/2021ApJ...919..107L}
}

@ARTICLE{Lorek2018,
       author = {{Lorek}, S. and {Lacerda}, P. and {Blum}, J.},
        title = "{Local growth of dust- and ice-mixed aggregates as cometary building blocks in the solar nebula}",
      journal = {\aap},
         year = 2018,
        month = mar,
       volume = {611},
          eid = {A18},
        pages = {A18},
          doi = {10.1051/0004-6361/201630175},
       adsurl = {https://ui.adsabs.harvard.edu/abs/2018A&A...611A..18L}
}

@ARTICLE{Lynden-BellPringle1974,
       author = {{Lynden-Bell}, D. and {Pringle}, J.~E.},
        title = "{The evolution of viscous discs and the origin of the nebular variables.}",
      journal = {\mnras},
         year = 1974,
        month = sep,
       volume = {168},
        pages = {603-637},
          doi = {10.1093/mnras/168.3.603},
       adsurl = {https://ui.adsabs.harvard.edu/abs/1974MNRAS.168..603L}
}

@ARTICLE{Mathis1977,
       author = {{Mathis}, J.~S. and {Rumpl}, W. and {Nordsieck}, K.~H.},
        title = "{The size distribution of interstellar grains.}",
      journal = {\apj},
         year = 1977,
        month = oct,
       volume = {217},
        pages = {425-433},
          doi = {10.1086/155591},
       adsurl = {https://ui.adsabs.harvard.edu/abs/1977ApJ...217..425M}
}

@ARTICLE{Michoulier2024,
       author = {{Michoulier}, St{\'e}phane and {Gonzalez}, Jean-Fran{\c{c}}ois and {Price}, Daniel J.},
        title = "{Compaction during fragmentation and bouncing produces realistic dust grain porosities in protoplanetary discs}",
      journal = {\aap},
         year = 2024,
        month = aug,
       volume = {688},
          eid = {A31},
        pages = {A31},
          doi = {10.1051/0004-6361/202449719},
archivePrefix = {arXiv},
       eprint = {2406.15622},
 primaryClass = {astro-ph.EP},
       adsurl = {https://ui.adsabs.harvard.edu/abs/2024A&A...688A..31M}
}

@ARTICLE{Okuzumi2009a,
author = {{Okuzumi}, Satoshi},
title = "{Electric Charging of Dust Aggregates and its Effect on Dust Coagulation in Protoplanetary Disks}",
journal = {\apj},
year = 2009,
month = jun,
volume = {698},
number = {2},
pages = {1122-1135},
doi = {10.1088/0004-637X/698/2/1122},
archivePrefix = {arXiv},
eprint = {0901.2886},
primaryClass = {astro-ph.EP},
adsurl = {ui.adsabs.harvard.edu/abs/2009ApJ...698.1122O}
}

@ARTICLE{Okuzumi2009b,
       author = {{Okuzumi}, Satoshi and {Tanaka}, Hidekazu and {Sakagami}, Masa-aki},
        title = "{Numerical Modeling of the Coagulation and Porosity Evolution of Dust Aggregates}",
      journal = {\apj},
         year = 2009,
        month = dec,
       volume = {707},
       number = {2},
        pages = {1247-1263},
          doi = {10.1088/0004-637X/707/2/1247},
archivePrefix = {arXiv},
       eprint = {0911.0239},
 primaryClass = {astro-ph.EP},
       adsurl = {https://ui.adsabs.harvard.edu/abs/2009ApJ...707.1247O}
}

@ARTICLE{Okuzumi2011,
       author = {{Okuzumi}, Satoshi and {Tanaka}, Hidekazu and {Takeuchi}, Taku and {Sakagami}, Masa-aki},
        title = "{Electrostatic Barrier Against Dust Growth in Protoplanetary Disks. I. Classifying the Evolution of Size Distribution}",
      journal = {\apj},
         year = 2011,
        month = apr,
       volume = {731},
       number = {2},
          eid = {95},
        pages = {95},
          doi = {10.1088/0004-637X/731/2/95},
archivePrefix = {arXiv},
       eprint = {1009.3199},
 primaryClass = {astro-ph.EP},
       adsurl = {https://ui.adsabs.harvard.edu/abs/2011ApJ...731...95O}
}

@ARTICLE{Okuzumi2012,
       author = {{Okuzumi}, Satoshi and {Tanaka}, Hidekazu and {Kobayashi}, Hiroshi and {Wada}, Koji},
        title = "{Rapid Coagulation of Porous Dust Aggregates outside the Snow Line: A Pathway to Successful Icy Planetesimal Formation}",
      journal = {\apj},
         year = 2012,
        month = jun,
       volume = {752},
       number = {2},
          eid = {106},
        pages = {106},
          doi = {10.1088/0004-637X/752/2/106},
archivePrefix = {arXiv},
       eprint = {1204.5035},
 primaryClass = {astro-ph.EP},
       adsurl = {https://ui.adsabs.harvard.edu/abs/2012ApJ...752..106O}
}

@ARTICLE{Ormel2007,
       author = {{Ormel}, C.~W. and {Spaans}, M. and {Tielens}, A.~G.~G.~M.},
        title = "{Dust coagulation in protoplanetary disks: porosity matters}",
      journal = {\aap},
         year = 2007,
        month = jan,
       volume = {461},
       number = {1},
        pages = {215-232},
          doi = {10.1051/0004-6361:20065949},
archivePrefix = {arXiv},
       eprint = {astro-ph/0610030},
 primaryClass = {astro-ph},
       adsurl = {https://ui.adsabs.harvard.edu/abs/2007A&A...461..215O}
}

@ARTICLE{OrmelCuzzi2007,
       author = {{Ormel}, C.~W. and {Cuzzi}, J.~N.},
        title = "{Closed-form expressions for particle relative velocities induced by turbulence}",
      journal = {\aap},
         year = 2007,
        month = may,
       volume = {466},
       number = {2},
        pages = {413-420},
          doi = {10.1051/0004-6361:20066899},
archivePrefix = {arXiv},
       eprint = {astro-ph/0702303},
 primaryClass = {astro-ph},
       adsurl = {https://ui.adsabs.harvard.edu/abs/2007A&A...466..413O}
}

@ARTICLE{Oshiro2025,
       author = {{Oshiro}, Haruto and {Tatsuuma}, Misako and {Okuzumi}, Satoshi and {Tanaka}, Hidekazu},
        title = "{Investigating the Bouncing Barrier with Collision Simulations of Compressed Dust Aggregates}",
      journal = {\apj},
         year = 2025,
        month = apr,
       volume = {983},
       number = {1},
          eid = {75},
        pages = {75},
          doi = {10.3847/1538-4357/adbf04},
archivePrefix = {arXiv},
       eprint = {2502.03107},
 primaryClass = {astro-ph.EP},
       adsurl = {https://ui.adsabs.harvard.edu/abs/2025ApJ...983...75O}
}

@ARTICLE{Schraepler2012,
       author = {{Schr{\"a}pler}, Rainer and {Blum}, J{\"u}rgen and {Seizinger}, Alexander and {Kley}, Wilhelm},
        title = "{The Physics of Protoplanetesimal Dust Agglomerates. VII. The Low-velocity Collision Behavior of Large Dust Agglomerates}",
      journal = {\apj},
         year = 2012,
        month = oct,
       volume = {758},
       number = {1},
          eid = {35},
        pages = {35},
          doi = {10.1088/0004-637X/758/1/35},
archivePrefix = {arXiv},
       eprint = {1208.3095},
 primaryClass = {astro-ph.EP},
       adsurl = {https://ui.adsabs.harvard.edu/abs/2012ApJ...758...35S}
}

@ARTICLE{Schraepler2022,
       author = {{Schr{\"a}pler}, Rainer R. and {Landeck}, Wolf A. and {Blum}, J{\"u}rgen},
        title = "{Collisional properties of cm-sized high-porosity ice and dust aggregates and their applications to early planet formation}",
      journal = {\mnras},
         year = 2022,
        month = feb,
       volume = {509},
       number = {4},
        pages = {5641-5656},
          doi = {10.1093/mnras/stab3348},
archivePrefix = {arXiv},
       eprint = {2111.09141},
 primaryClass = {astro-ph.EP},
       adsurl = {https://ui.adsabs.harvard.edu/abs/2022MNRAS.509.5641S}
}

@ARTICLE{SeizingerKley2013,
       author = {{Seizinger}, A. and {Kley}, W.},
        title = "{Bouncing behavior of microscopic dust aggregates}",
      journal = {\aap},
         year = 2013,
        month = mar,
       volume = {551},
          eid = {A65},
        pages = {A65},
          doi = {10.1051/0004-6361/201220946},
archivePrefix = {arXiv},
       eprint = {1301.3629},
 primaryClass = {astro-ph.EP},
       adsurl = {https://ui.adsabs.harvard.edu/abs/2013A&A...551A..65S}
}

@ARTICLE{ShakuraSunyaev1973,
       author = {{Shakura}, N.~I. and {Sunyaev}, R.~A.},
        title = "{Black holes in binary systems. Observational appearance.}",
      journal = {\aap},
         year = 1973,
        month = jan,
       volume = {24},
        pages = {337-355},
       adsurl = {https://ui.adsabs.harvard.edu/abs/1973A&A....24..337S}
}

@ARTICLE{Smoluchowski1916,
author = {{Smoluchowski}, M.~V.},
title = "{Drei Vortrage uber Diffusion, Brownsche Bewegung und Koagulation von Kolloidteilchen}",
journal = {Zeitschrift fur Physik},
year = 1916,
month = jan,
volume = {17},
pages = {557-585},
adsurl = {ui.adsabs.harvard.edu/abs/1916ZPhy...17..557S}
}

@ARTICLE{StammlerBirnstiel2022,
       author = {{Stammler}, Sebastian M. and {Birnstiel}, Tilman},
        title = "{DustPy: A Python Package for Dust Evolution in Protoplanetary Disks}",
      journal = {\apj},
         year = 2022,
        month = aug,
       volume = {935},
       number = {1},
          eid = {35},
        pages = {35},
          doi = {10.3847/1538-4357/ac7d58},
archivePrefix = {arXiv},
       eprint = {2207.00322},
 primaryClass = {astro-ph.EP},
       adsurl = {https://ui.adsabs.harvard.edu/abs/2022ApJ...935...35S}
}

@ARTICLE{Suyama2008,
       author = {{Suyama}, Toru and {Wada}, Koji and {Tanaka}, Hidekazu},
        title = "{Numerical Simulation of Density Evolution of Dust Aggregates in Protoplanetary Disks. I. Head-on Collisions}",
      journal = {\apj},
         year = 2008,
        month = sep,
       volume = {684},
       number = {2},
        pages = {1310-1322},
          doi = {10.1086/590143},
       adsurl = {https://ui.adsabs.harvard.edu/abs/2008ApJ...684.1310S}
}

@ARTICLE{Tanaka2005,
author = {{Tanaka}, Hidekazu and {Himeno}, Youhei and {Ida}, Shigeru},
title = "{Dust Growth and Settling in Protoplanetary Disks and Disk Spectral Energy Distributions. I. Laminar Disks}",
journal = {\apj},
year = 2005,
month = may,
volume = {625},
number = {1},
pages = {414-426},
doi = {10.1086/429658},
archivePrefix = {arXiv},
eprint = {astro-ph/0502287},
primaryClass = {astro-ph},
adsurl = {ui.adsabs.harvard.edu/abs/2005ApJ...625..414T}
}

@ARTICLE{Tazaki2021,
       author = {{Tazaki}, Ryo},
        title = "{Analytic expressions for geometric cross-sections of fractal dust aggregates}",
      journal = {\mnras},
         year = 2021,
        month = jun,
       volume = {504},
       number = {2},
        pages = {2811-2821},
          doi = {10.1093/mnras/stab1069},
archivePrefix = {arXiv},
       eprint = {2104.06804},
 primaryClass = {astro-ph.EP},
       adsurl = {https://ui.adsabs.harvard.edu/abs/2021MNRAS.504.2811T}
}

@ARTICLE{Wada2007,
       author = {{Wada}, Koji and {Tanaka}, Hidekazu and {Suyama}, Toru and {Kimura}, Hiroshi and {Yamamoto}, Tetsuo},
        title = "{Numerical Simulation of Dust Aggregate Collisions. I. Compression and Disruption of Two-Dimensional Aggregates}",
      journal = {\apj},
         year = 2007,
        month = may,
       volume = {661},
       number = {1},
        pages = {320-333},
          doi = {10.1086/514332},
       adsurl = {https://ui.adsabs.harvard.edu/abs/2007ApJ...661..320W}
}

@ARTICLE{Wada2008,
       author = {{Wada}, Koji and {Tanaka}, Hidekazu and {Suyama}, Toru and {Kimura}, Hiroshi and {Yamamoto}, Tetsuo},
        title = "{Numerical Simulation of Dust Aggregate Collisions. II. Compression and Disruption of Three-Dimensional Aggregates in Head-on Collisions}",
      journal = {\apj},
         year = 2008,
        month = apr,
       volume = {677},
       number = {2},
        pages = {1296-1308},
          doi = {10.1086/529511},
       adsurl = {https://ui.adsabs.harvard.edu/abs/2008ApJ...677.1296W}
}

@ARTICLE{Wada2009,
       author = {{Wada}, Koji and {Tanaka}, Hidekazu and {Suyama}, Toru and {Kimura}, Hiroshi and {Yamamoto}, Tetsuo},
        title = "{Collisional Growth Conditions for Dust Aggregates}",
      journal = {\apj},
         year = 2009,
        month = sep,
       volume = {702},
       number = {2},
        pages = {1490-1501},
          doi = {10.1088/0004-637X/702/2/1490},
       adsurl = {https://ui.adsabs.harvard.edu/abs/2009ApJ...702.1490W}
}

@ARTICLE{Wada2011,
       author = {{Wada}, Koji and {Tanaka}, Hidekazu and {Suyama}, Toru and {Kimura}, Hiroshi and {Yamamoto}, Tetsuo},
        title = "{The Rebound Condition of Dust Aggregates Revealed by Numerical Simulation of Their Collisions}",
      journal = {\apj},
         year = 2011,
        month = aug,
       volume = {737},
       number = {1},
          eid = {36},
        pages = {36},
          doi = {10.1088/0004-637X/737/1/36},
       adsurl = {https://ui.adsabs.harvard.edu/abs/2011ApJ...737...36W}
}

@ARTICLE{Weidenschilling1977,
       author = {{Weidenschilling}, S.~J.},
        title = "{Aerodynamics of solid bodies in the solar nebula.}",
      journal = {\mnras},
         year = 1977,
        month = jul,
       volume = {180},
        pages = {57-70},
          doi = {10.1093/mnras/180.2.57},
       adsurl = {https://ui.adsabs.harvard.edu/abs/1977MNRAS.180...57W}
}

@ARTICLE{Weidenschilling1997,
author = {{Weidenschilling}, S.~J.},
title = "{The Origin of Comets in the Solar Nebula: A Unified Model}",
journal = {\icarus},
year = 1997,
month = jun,
volume = {127},
number = {2},
pages = {290-306},
doi = {10.1006/icar.1997.5712},
adsurl = {ui.adsabs.harvard.edu/abs/1997Icar..127..290W}
}

@ARTICLE{Xiang2020,
       author = {{Xiang}, C. and {Matthews}, L.~S. and {Carballido}, A. and {Hyde}, T.~W.},
        title = "{Detailed Model of the Growth of Fluffy Dust Aggregates in a Protoplanetary Disk: Effects of Nebular Conditions}",
      journal = {\apj},
         year = 2020,
        month = jul,
       volume = {897},
       number = {2},
          eid = {182},
        pages = {182},
          doi = {10.3847/1538-4357/ab96c2},
archivePrefix = {arXiv},
       eprint = {1911.04589},
 primaryClass = {astro-ph.EP},
       adsurl = {https://ui.adsabs.harvard.edu/abs/2020ApJ...897..182X}
}

@ARTICLE{Yang2017,
       author = {{Yang}, Chao-Chin and {Johansen}, Anders and {Carrera}, Daniel},
        title = "{Concentrating small particles in protoplanetary disks through the streaming instability}",
      journal = {\aap},
         year = 2017,
        month = oct,
       volume = {606},
          eid = {A80},
        pages = {A80},
          doi = {10.1051/0004-6361/201630106},
archivePrefix = {arXiv},
       eprint = {1611.07014},
 primaryClass = {astro-ph.EP},
       adsurl = {https://ui.adsabs.harvard.edu/abs/2017A&A...606A..80Y}
}

@ARTICLE{YoudinGoodman2005,
       author = {{Youdin}, Andrew N. and {Goodman}, Jeremy},
        title = "{Streaming Instabilities in Protoplanetary Disks}",
      journal = {\apj},
         year = 2005,
        month = feb,
       volume = {620},
       number = {1},
        pages = {459-469},
          doi = {10.1086/426895},
archivePrefix = {arXiv},
       eprint = {astro-ph/0409263},
 primaryClass = {astro-ph},
       adsurl = {https://ui.adsabs.harvard.edu/abs/2005ApJ...620..459Y}
}

@ARTICLE{ZsomDullemond2008,
author = {{Zsom}, A. and {Dullemond}, C.~P.},
title = "{A representative particle approach to coagulation and fragmentation of dust aggregates and fluid droplets}",
journal = {\aap},
year = 2008,
month = oct,
volume = {489},
number = {2},
pages = {931-941},
doi = {10.1051/0004-6361:200809921},
archivePrefix = {arXiv},
eprint = {0807.5052},
primaryClass = {astro-ph},
adsurl = {ui.adsabs.harvard.edu/abs/2008A&A...489..931Z}
}

@ARTICLE{Zsom2010,
       author = {{Zsom}, A. and {Ormel}, C.~W. and {G{\"u}ttler}, C. and {Blum}, J. and {Dullemond}, C.~P.},
        title = "{The outcome of protoplanetary dust growth: pebbles, boulders, or planetesimals? II. Introducing the bouncing barrier}",
      journal = {\aap},
         year = 2010,
        month = apr,
       volume = {513},
          eid = {A57},
        pages = {A57},
          doi = {10.1051/0004-6361/200912976},
archivePrefix = {arXiv},
       eprint = {1001.0488},
 primaryClass = {astro-ph.EP},
       adsurl = {https://ui.adsabs.harvard.edu/abs/2010A&A...513A..57Z}
}

\begin{appendix}
\section{Area dispersion}
\label{sec:appendix_area_dispersion}
Two fractal particles of the same mass can have slightly different cross sections due to the statistical nature of their structure. We refer to this as area dispersion. To take into account the effect of area dispersion, we adjusted the expressions for relative velocities. In the following, we derive these adjusted expressions (Sect. \ref{sec:area_dispersion_derivation}) and then investigate the effect of area dispersion (Sect. \ref{sec:discussion_area_dispersion}).
\subsection{Derivation of relative velocity expressions}
\label{sec:area_dispersion_derivation}
Here, we derive the expressions for the relative velocities including area dispersion. We assume that for each mass bin, the mass-to-area ratio \(B=m/A\) (and therefore the Stokes number) is dispersed following a Log-Gauss-distribution:
\begin{equation}
	p(B)\mathrm{d}\ln B = \frac{1}{\sqrt{2\pi}\sigma}\exp\left[-\frac{\left(\ln\left(B/\mu^*\right)\right)^2}{2\sigma^2}\right]\mathrm{d}\ln B,
	\label{eq:loggauss}
\end{equation}
where \(\mu^* \!=\! \exp(\mu)\) and \(\mu \!=\! \overline{\ln B}\) is the mean of \(\ln B\) and \(\sigma\) is the standard deviation of \(\ln B\). The dispersion of mass-to-area ratio leads to a dispersion of Stokes numbers. We include area dispersion into the relative velocity equations by feeding them with the distribution of Stokes numbers and then taking the root mean square.

We note that 
\begin{equation}
	\overline{B^x} = \overline{B}^x\exp\left(\frac{x^2-x}{2}\sigma^2\right),
	\label{eq:Bx}
\end{equation}
for an arbitrary exponent \(x\).
We assume that the mass-to-area ratios are not correlated between two particles, therefore \(\overline{B_1B_2} = \overline{B}_1\overline{B}_2\) holds.

Although we assume a Log-Gauss distribution, the results for radial drift, turbulence and, for \(\mathrm{St}\ll \alpha\), vertical settling are more general. In this case, it is sufficient to assume that the variance can be written as 
\begin{equation}
\text{Var}\left(B\right) = \overline{B}^2\epsilon^2.
\end{equation}
The reason is that \(\overline{B^2}\) can be written as \(\overline{B}^2 + \mathrm{Var}\left(B\right)\). Thus, it is not necessary to assume a particular distribution, as long as the mean and the variance are known. However, the results for azimuthal drift and the \(\alpha\lesssim\mathrm{St}\) regime of vertical settling are less general, because higher powers of \(\mathrm{St}\) appear, whose averages as a function of \(\overline{\mathrm{St}}\) depend on the chosen distribution. For the Log-Gauss distribution, \(\epsilon\) is given by
\begin{equation}
	\epsilon^2 = \exp\left(\sigma^2\right) - 1.
\end{equation}
To keep it general, we distinguish between the width parameter \(\epsilon_1\) and \(\epsilon_2\) for two different mass bins in the derivation. For our models, however, we assume that \(\epsilon\) is the same for all mass bins.

\subsubsection{Radial drift}
The default expression for radial drift in \texttt{DustPy} is given by \citep{StammlerBirnstiel2022}
\begin{equation}
	v_\mathrm{rel}^\mathrm{rad} = \left|\frac{v_\mathrm{g} + 2v_\mathrm{drif}^\mathrm{max}\mathrm{St}_1}{1+\mathrm{St}_1^2} - \frac{v_\mathrm{g} + 2v_\mathrm{drif}^\mathrm{max}\mathrm{St}_2}{1+\mathrm{St}_2^2}\right|,
	\label{eq:rad_drift}
\end{equation}
where \(v_\mathrm{g}\) is the velocity of the gas, and
\begin{equation}
	v_\mathrm{drift}^\mathrm{max} = \frac{1}{2} \left(\frac{H_P}{r}\right)^2 \frac{\partial\ln P}{\partial\ln r}.
\end{equation}
We approximate 
\begin{equation*}
	\frac{\mathrm{St}}{1+\mathrm{St}^2}\approx\frac{\mathrm{St}}{1+\overline{\mathrm{St}^2}} = \frac{\mathrm{St}}{1+\left(1+\epsilon^2\right)\overline{\mathrm{St}}^2}.
\end{equation*}
The mean square is then given by
\begin{align}
	\nonumber
	\overline{\left(v_\mathrm{rel}^\text{rad}\right)^2} &=
	\left(\frac{v_\mathrm{g} + 2v_\text{drift}^\text{max}\overline{\mathrm{St}}_1}{1+\left(1+\epsilon_1^2\right)\overline{\mathrm{St}}_1^2} - 
	\frac{v_\mathrm{g} + 2v_\text{drift}^\text{max}\overline{\mathrm{St}}_2}{1+\left(1+\epsilon_2^2\right)\overline{\mathrm{St}}_2^2}\right)^2 \\ &+
	\epsilon_1^2\left(\frac{2v_\text{drift}^\text{max}\overline{\mathrm{St}}_1}{1+\left(1+\epsilon_1^2\right)\overline{\mathrm{St}}_1^2}\right)^2 + 
	\epsilon_2^2\left(\frac{2v_\text{drift}^\text{max}\overline{\mathrm{St}}_2}{1+\left(1+\epsilon_2^2\right)\overline{\mathrm{St}}_2^2}\right)^2.
\label{eq:rad_approx}
\end{align}
The first term in Eq. \eqref{eq:rad_approx} looks similar to the original expression but there are additional terms which depend on the dispersion parameter \(\epsilon\).

\subsubsection{Azimuthal drift}
The default expression for relative velocities induced by azimuthal drift is given by \citep{StammlerBirnstiel2022}
\begin{equation}
	v_\mathrm{rel}^\mathrm{azi} = \left|v_\text{drift}^{\text{max}}\frac{\mathrm{St}_2^2 - \mathrm{St}_1^2}{\left(1+\text{St}_1^2\right)\left(1+\text{St}_2^2\right)}\right|.
	\label{eq:azi_drift}
\end{equation}
We approximate
\begin{equation*}
	\left|\frac{\mathrm{St}^2_1 - \mathrm{St}^2_2}{\left(1+\mathrm{St}_1^2\right)\left(1+\mathrm{St}_2^2\right)}\right| \approx \left|\frac{\mathrm{St}^2_1 - \mathrm{St}^2_2}{\left(1+\overline{\mathrm{St}_1^2}\right)\left(1+\overline{\mathrm{St}_2^2}\right)}\right|.
\end{equation*}
The mean square of Eq. \eqref{eq:azi_drift} is then given by
\begin{align}
	\nonumber\overline{\left(v_\mathrm{rel}^\mathrm{azi}\right)^2} &= v_\mathrm{drift}^\mathrm{max}
	\left(\frac{1}{1+\left(1+\epsilon_1^2\right)\overline{\mathrm{St}}_1^2} - \frac{1}{1+\left(1+\epsilon_2^2\right)\overline{\mathrm{St}}_2^2}\right)^2\\ &+
	5v_\mathrm{drift}^\mathrm{max}\frac{\epsilon_1^2\overline{\mathrm{St}}_1^4 
	+ \epsilon_2^2\overline{\mathrm{St}}_2^4}
	{\left(1+\left(1+\epsilon_1^2\right)\overline{\mathrm{St}}_1^2\right)^2\left(1+\left(1+\epsilon_2^2\right)\overline{\mathrm{St}}_2^2\right)^2},
	\label{eq:azi_approx}
\end{align}
where we neglected \(\mathcal{O}\left(\epsilon^4\right)\) terms.

\subsubsection{Vertical settling}
The default expression for vertical settling is given by \citep{StammlerBirnstiel2022}
\begin{equation}
	v_\mathrm{rel}^\mathrm{sett} = c_\mathrm{s,iso} \sqrt{\alpha} 
	\left|\frac{\min\left(\mathrm{St}_1, 0.5\right)}{\sqrt{\alpha + \mathrm{St}_1}} - \frac{\min\left(\mathrm{St}_2, 0.5\right)}{\sqrt{\alpha + \mathrm{St}_2}}\right|,
	\label{eq:vert_sett}
\end{equation}
where \(c_\mathrm{s,iso} = c_\mathrm{s} / \sqrt{\gamma}\), is the isothermal sound speed. 

First, we consider the case, where \(\mathrm{St}_1, \mathrm{St}_2 < 0.5\). 
For \(\mathrm{St} \ll \alpha\) we again neglect the area dispersion of the denominator:
\begin{equation*}
	\frac{\mathrm{St}}{\sqrt{\alpha+\mathrm{St}}} \approx \frac{\mathrm{St}}{\sqrt{\alpha+\overline{\mathrm{St}}}}.
\end{equation*}
For \(\mathrm{St} \gg \alpha\), we rewrite the term
\begin{equation*}
	\frac{\mathrm{St}}{\sqrt{\alpha + \mathrm{St}}} =\frac{\sqrt{\mathrm{St}}}{\sqrt{1+\alpha/\mathrm{St}}},
\end{equation*}
where we can then again neglect the area dispersion of the denominator. 

For \(\mathrm{St}_1, \mathrm{St}_2 \ll \alpha\), the mean square of Eq. \eqref{eq:vert_sett} is given by
\begin{align}
	\nonumber\overline{\left(v_\mathrm{rel}^\mathrm{sett}\right)^2} 
	&= \alpha c_\mathrm{s,iso}^2 \left(\frac{\overline{\mathrm{St}_1}}{\sqrt{\alpha+\overline{\mathrm{St}_1}}}
	- \frac{\overline{\mathrm{St}_2}}{\sqrt{\alpha+\overline{\mathrm{St}_2}}}\right)^2 \\ &+
	\alpha c_\mathrm{s,iso}^2  \epsilon_1^2 \left(\frac{\overline{\mathrm{St}_1}}{\sqrt{\alpha+\overline{\mathrm{St}_1}}}\right)^2 + 
	\alpha c_\mathrm{s,iso}^2 \epsilon_2^2 \left(\frac{\overline{\mathrm{St}_2}}{\sqrt{\alpha+\overline{\mathrm{St}_2}}}\right)^2.
	\label{eq:v_vert_low_alpha}
\end{align}
For \(\mathrm{St}_2 \ll \alpha \ll \mathrm{St}_1\), the mean square of Eq. \eqref{eq:vert_sett} is given by
\begin{align}
	\nonumber\overline{\left(v_\mathrm{rel}^\mathrm{sett}\right)^2} &=
	\alpha c_\mathrm{s,iso}^2 \left(\frac{\overline{\mathrm{St}}_1}{\sqrt{\alpha + \overline{\mathrm{St}}_1}}
	- \frac{\overline{\mathrm{St}}_2}{\sqrt{\alpha + \overline{\mathrm{St}}_2}}\right)^2 \\ &+ 
	\alpha c_\mathrm{s,iso}^2\frac{\epsilon_1^2}{4}\frac{\overline{\mathrm{St}}_1}{\sqrt{\alpha+\overline{\mathrm{St}}_1}}
	\frac{\overline{\mathrm{St}}_2}{\sqrt{\alpha+\overline{\mathrm{St}}_2}} +
	\alpha c_\mathrm{s,iso}^2 \epsilon_2^2\frac{\overline{\mathrm{St}}_2^2}{\alpha + \mathrm{St}_2},
	\label{eq:vert_2}
\end{align}
and vice versa for \(\mathrm{St}_1 \ll \alpha \ll \mathrm{St}_2\). For \(\alpha \ll \mathrm{St}_1, \mathrm{St}_2\), the mean square of Eq. \eqref{eq:vert_sett} is given by
\begin{align}
	\nonumber\overline{\left(v_\mathrm{rel}^\mathrm{sett}\right)^2} &=
	\alpha c_\mathrm{s,iso}^2\left(\frac{\overline{\mathrm{St}}_1}{\sqrt{\alpha + \overline{\mathrm{St}}_1}}
	- \frac{\overline{\mathrm{St}}_2}{\sqrt{\alpha + \overline{\mathrm{St}}_2}}\right)^2\\ &+
	 \frac{1}{4} \alpha c_\mathrm{s}^2\left(\epsilon_1^2+\epsilon_2^2\right)\frac{\overline{\mathrm{St}}_1}{\sqrt{\alpha+\overline{\mathrm{St}}_1}}
	\frac{\overline{\mathrm{St}}_2}{\sqrt{\alpha+\overline{\mathrm{St}}_2}}.
	\label{eq:vert_3}
\end{align}
To merge Eqs. \eqref{eq:vert_sett}, \eqref{eq:vert_2}, and \eqref{eq:vert_3} together, we define coefficients
\begin{equation*}
	p = \frac{\alpha}{\overline{\mathrm{St}} + \alpha}\quad\text{and}\quad q = 1 - p.
\end{equation*}
Furthermore, to include the case \(\mathrm{St}>0.5\), we define coefficients 
\begin{equation*}
	y = \begin{cases}
	1 \text{ for } \mathrm{St}\leq0.5,\\
	0\text{ for } \mathrm{St}>0.5,
	\end{cases}
\end{equation*}
and \(z = 1 - y\). 
Putting everything together yields
\begin{align}	
	\nonumber\overline{\left(v_\mathrm{rel}^\mathrm{sett}\right)^2} &= \nonumber \alpha c_\mathrm{s,iso}^2
	\left[\ \left(\frac{\min\left(\overline{\mathrm{St}}_1, 0.5\right)}{\sqrt{\alpha+\overline{\mathrm{St}}_1}} 
	- \frac{\min\left(\overline{\mathrm{St}}_2, 0.5\right)}{\sqrt{\alpha+\overline{\mathrm{St}}_2}}\right)^2 \right.\\
	\nonumber &+ y_1 p_1 \epsilon_1^2 
	\frac{\overline{\mathrm{St}}_1^2}{\alpha+\overline{\mathrm{St}}_1} +
	y_2 p_2 \epsilon_2^2 
	\frac{\overline{\mathrm{St}}_2^2}{\alpha+\overline{\mathrm{St}}_2}\\
	\nonumber &+ \left(y_1q_1\frac{\epsilon_1^2}{4}+y_2q_2\frac{\epsilon_2^2}{4}\right)
	\frac{\min\left(\overline{\mathrm{St}}_1,0.5\right)
	\min\left(\overline{\mathrm{St}}_2, 0.5\right)}{\sqrt{\left(\alpha+\overline{\mathrm{St}}_1\right)
	\left(\alpha+\overline{\mathrm{St}}_2\right)}}\\
	\nonumber&+ z_1\epsilon_1^2\left(\frac{0.25}{\alpha+\mathrm{St}_1} 
	- \frac{3}{4} \frac{0.5\min\left(\overline{\mathrm{St}}_2, 0.5\right)}
	{\sqrt{\left(\alpha+\overline{\mathrm{St}}_1\right)\left(\alpha+\overline{\mathrm{St}}_2\right)}}\right)\\
	&+z_2\epsilon_2^2\left.\left(\frac{0.25}{\alpha+\mathrm{St}_2}
	- \frac{3}{4} \frac{0.5\min\left(\overline{\mathrm{St}}_1, 0.5\right)}
	{\sqrt{\left(\alpha+\overline{\mathrm{St}}_2\right)\left(\alpha+\overline{\mathrm{St}}_1\right)}}\right)\ \right],
	\label{eq:sett_approx}
\end{align}
where we assumed \(\overline{\min\left(\mathrm{St}, 0.5\right)}=\min\left(\overline{\mathrm{St}}, 0.5\right)\). This is not always true but the deviations are small because \(\epsilon\ll1\).

The total relative velocity is given by the quadratic sum of the contributions:
\begin{equation}
	\label{eq:v_rel_tot}
	v_\mathrm{rel}^\mathrm{tot} = \sqrt{\left(v_\mathrm{rel}^\mathrm{brown}\right)^2 +
											\left(v_\mathrm{rel}^\mathrm{sett}\right)^2 + 
											\left(v_\mathrm{rel}^\mathrm{rad}\right)^2 +
											\left(v_\mathrm{rel}^\mathrm{azi}\right)^2 +
											\left(v_\mathrm{rel}^\mathrm{turb}\right)^2}.
\end{equation}

\subsubsection{Turbulence}
For turbulence, \texttt{DustPy} differentiates between different turbulence regimes, based on the stopping time \(t_\mathrm{L}\) of the larger particle. Only the first two regimes depend on the difference between Stokes numbers. The other regimes therefore do not have to be modified.

For \(t_\mathrm{L} < 0.2 t_\eta\), the default equation is given by
\begin{equation}
	v_\mathrm{rel}^\text{turb} = \sqrt{\frac{3\alpha}{2}} \mathrm{Re}^{1/4} c_\mathrm{s}\left|\mathrm{St_1}-\mathrm{St_2}\right|.
	\label{eq:turb1}
\end{equation}
The mean square of Eq. \eqref{eq:turb1} is given by
\begin{equation}
	\overline{\left(v_\mathrm{rel}^\mathrm{turb}\right)^2} =
    \frac{3\alpha}{2} \mathrm{Re}^{1/2} c_\mathrm{s}^2 \left[
	 \left(\overline{\mathrm{St}}_1 - \overline{\mathrm{St}}_2\right)^2 
	 + \epsilon_1^2\overline{\mathrm{St}}_1^2 + \epsilon_2^2\overline{\mathrm{St}}_2^2\right].
	 \label{eq:turb1a}
\end{equation}

For \(0.2t_\eta\leq t_L < t_\eta / y_\mathrm{a}\), where \(y_\mathrm{a} \!=\! 1.6\), the turbulent relative velocity is given by
\begin{equation}
	v_\mathrm{rel}^\text{turb} = 
	\sqrt{\frac{3\alpha}{2}}c_\mathrm{s}\left[\frac{\mathrm{St_L} - \mathrm{St_S}}{\mathrm{St_L}+\mathrm{St_S}}
	\left(\frac{\mathrm{St_L}^2}{\mathrm{St_L} + \mathrm{Re}^{-1/2}} - \frac{\mathrm{St_S}^2}{\mathrm{St_S} + \mathrm{Re}^{-1/2}}\right)\right]^{1/2},
	\label{eq:turb2}
\end{equation}
where \(\mathrm{St_L}\) and \(\mathrm{St_S}\) are the larger and smaller Stokes number, respectively.
Since \(t_\mathrm{L} < t_\eta / y_\mathrm{a}\), we know that \(\mathrm{St_S}\leq\mathrm{St_L} < \mathrm{Re}^{-1/2}\). Therefore, the denominator of the terms inside the parentheses only weakly depends on the Stokes numbers. We approximate
\begin{equation*}
	\frac{1}{\mathrm{St_L} + \mathrm{Re}^{-1/2}} \approx \frac{1}{\overline{\mathrm{St_{LS}}} + \mathrm{Re}^{-1/2}},
\end{equation*}
and 
\begin{equation*}
	\frac{1}{\mathrm{St_S} + \mathrm{Re}^{-1/2}} \approx \frac{1}{\overline{\mathrm{St_{LS}}} + \mathrm{Re}^{-1/2}},
\end{equation*}
where where \(\overline{\mathrm{St_{LS}}} = 0.5 \left(\overline{\mathrm{St_L}} + \overline{\mathrm{St_S}}\right)\).

Then, the mean square of Eq. \eqref{eq:turb2} is given by
\begin{align}
	\overline{\left(v_\mathrm{rel}^\mathrm{turb}\right)^2} &=
	\sqrt{\frac{3\alpha}{2}}\frac{c_\mathrm{s}}{\overline{\mathrm{St_{SL}}} + \mathrm{Re}^{-1/2}}
	\left[\left(\overline{\mathrm{St_1}} - \overline{\mathrm{St_2}}\right)^2 +
	\overline{\mathrm{St_1}}^2\epsilon_1^2 + \overline{\mathrm{St_2}}^2\epsilon_2^2\right].
\end{align}

\subsection{The effect of area dispersion}
\label{sec:discussion_area_dispersion}
The models presented in Sects. \ref{sec:sc_models} and \ref{sec:ac_models} all included area dispersion. In the following, we want to discuss under which conditions area dispersion is important and when it can be neglected. We expect that area dispersion is important when the Stokes number only weakly depends on mass.
To verify this hypothesis, we ran additional models without including area dispersion, i.e., with the default expressions for relative velocities in \texttt{DustPy} \citep[see][]{StammlerBirnstiel2022}.

We explored SC models with \(D \!=\! 2\), \(D=2.05\), \(D \!=\! 2.5\) and \(D \!=\! 3\). The \(D \!=\! 2.05\) model is to investigate the sensitivity of the dust distribution, when the fractal dimension is slightly different from the special case of \(D \!=\! 2\) (where \(\mathrm{St} \!=\! \mathrm{const.}\)), for both the models with and without area dispersion. 

\begin{figure*}
   \centering
   \includegraphics[width=\hsize]{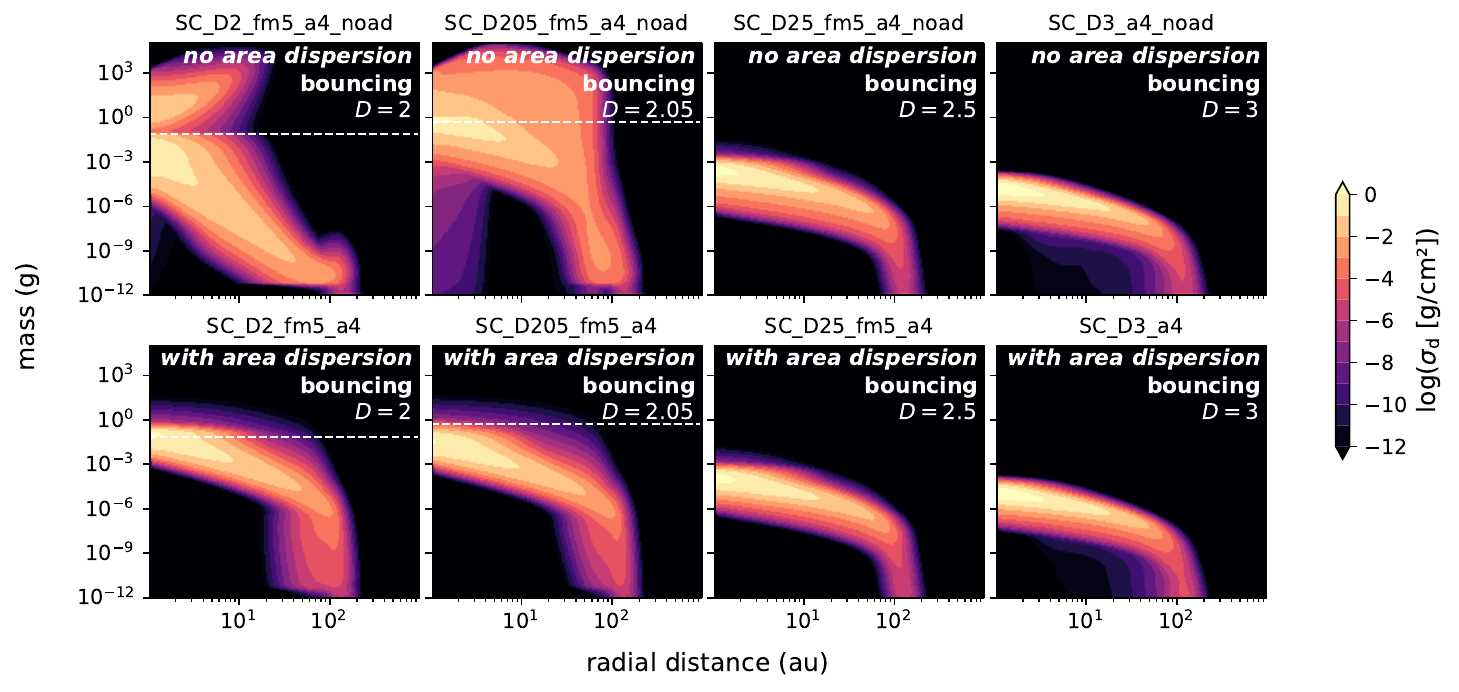}
      \caption{Dust distribution of SC models after \(10^6\) without area dispersion (top row) and with area dispersion (bottom row). The white dashed lines indicate the transition from fractal to non-fractal growth.}
         \label{fig:sc_compare_ad}
\end{figure*}

Since the Stokes number does not depend on mass for \(D=2\), we expect the area dispersion to have a large effect. Indeed, the dust distribution after \(10^6\) years of the SC model with \(D=2\) looks very different when area dispersion is not included (see Fig. \ref{fig:sc_compare_ad}). Also, the shape of the dust distribution for \(D \!=\! 2\) is very sensitive to the fractal dimension. Decreasing \(D\) only slightly to \(D \!=\! 2.05\) changes the dust distribution drastically. When area dispersion is included, this slight change of fractal dimension only leads to a slight change of the dust distribution. For \(D \!=\! 2.5\) and for \(D \!=\! 3\), there is no considerable difference between the models with and without area dispersion. 

\begin{figure}
   \centering
   \includegraphics[width=\hsize]{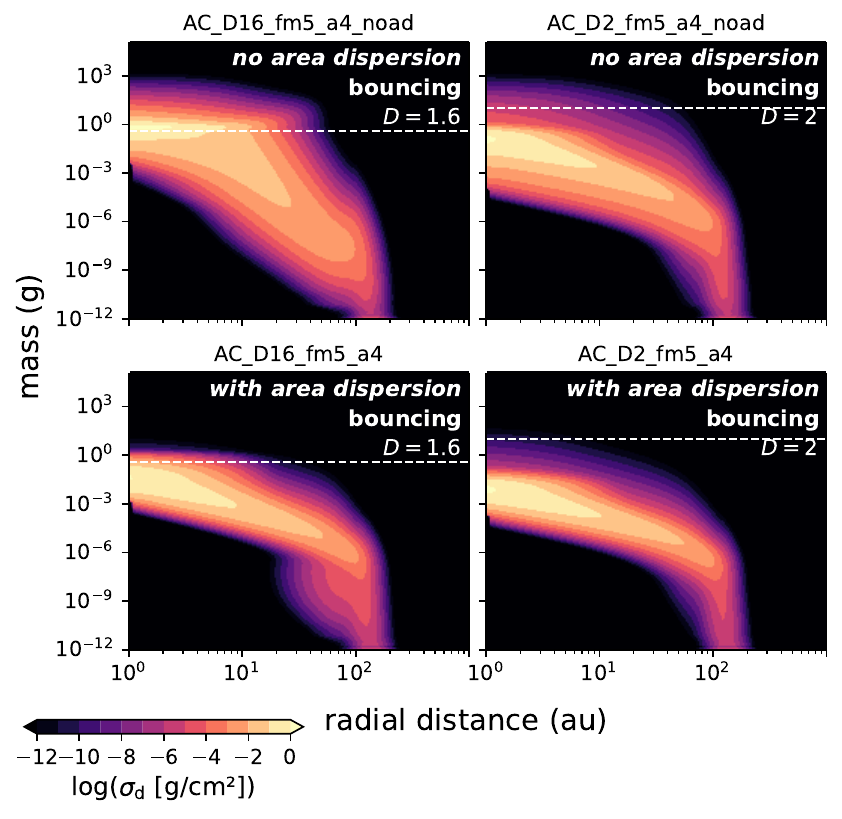}
      \caption{Same as Fig. \ref{fig:sc_compare_ad}, but for the AC models.}
         \label{fig:ac_compare_ad}
\end{figure}

For the AC models, the Stokes number is never truly independent of mass. However, for \(D \!\leq\! 2\) it approaches a constant value (see Fig. \ref{fig:ac_vs_sc_cross_sections}).  To investigate the importance of area dispersion in the case of advanced cross sections, we therefore ran AC models without including area dispersion for \(D=1.6\) and \(D=2\). We compare their dust distributions to those of the models with area dispersion in Fig. \ref{fig:ac_compare_ad}. For \(D \!=\! 2\), the difference between the models with and the models without area dispersion is visible, although less pronounced than for the SC models. This is because the Stokes number is not fully independent of mass. For \(D \!=\! 1.6\), the difference is larger because the mass dependence of the Stokes number is weaker.

\begin{figure*}
   \centering
   \includegraphics[width=\hsize]{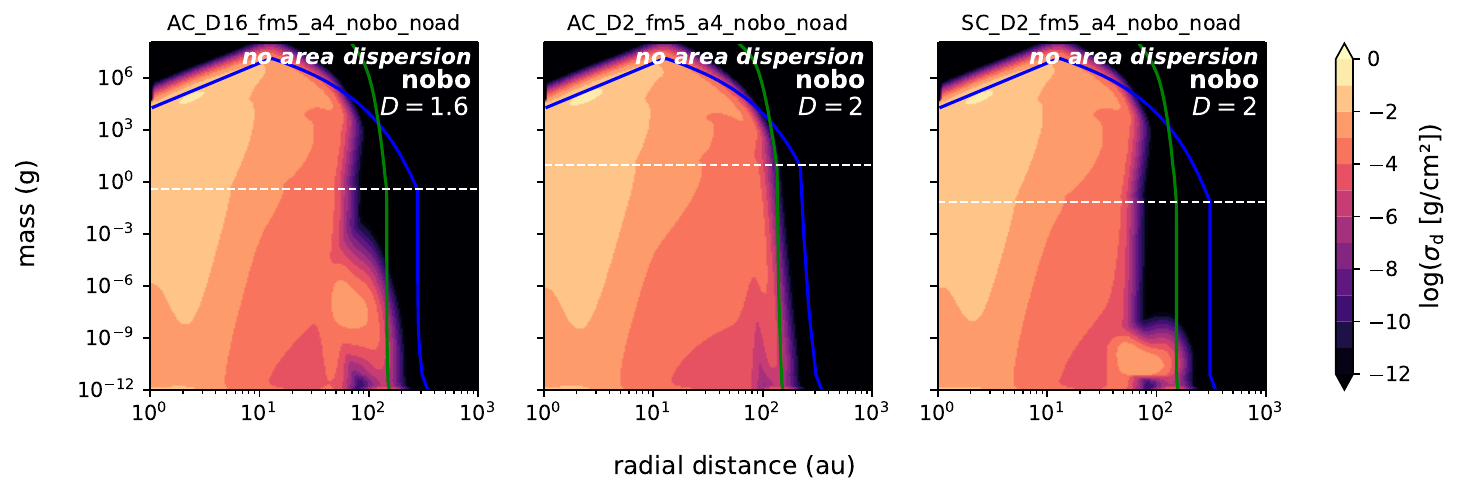}
      \caption{Dust distribution after \(10^6\) years of the SC and AC models without bouncing and without area dispersion. The colored lines indicate the growth barriers as in previous figures. The white dashed lines indicate the transition from fractal to non-fractal growth.}
         \label{fig:sigma_d_nobo_noad}
\end{figure*}

In the case without bouncing, the difference between models with and without area dispersion is only small even for \(D \!=\! 2\) for both SC and AC models. The reason is that the mass distribution is shaped by the fragmentation barrier, which lies in the non-fractal growth regime, where area dispersion is not important. Except for the very outer parts of the disk, where the particles in the models without area dispersion have not reached the fragmentation barrier yet (see Fig. \ref{fig:sigma_d_nobo_noad}), because growth is slower without area dispersion.

\section{Code quantities and diagnostics}
\label{sec:appendix_code_quantities}
\subsection{Dust distribution}
The dust distribution  \(\sigma_\mathrm{d}(r,m)\) is defined as 
\begin{equation}
	\Delta\Sigma_\mathrm{d} = \int_{\log m_1}^{\log m_2}\sigma_\mathrm{d}(r,m)\mathrm{d}\log m,
	\label{eq:def_sigma_d}
\end{equation}
where \(\Delta\Sigma_\mathrm{d}\) is the surface density of dust particles with masses between \(m_1\) and \(m_2\). 

Internally, \texttt{DustPy} works with the quantity \citep{StammlerBirnstiel2022}\footnote{see \texttt{DustPy} documentation: \url{https://stammler.github.io/dustpy/}}
\begin{equation}
	\Sigma_{\text{d},j}(r) = 
	\int_{m_{j-1/2}}^{m_{j+1/2}}\sigma_\text{d}(r, m)\frac{\mathrm{d}m}{m} ,
	\label{eq:Sigma_ij}
\end{equation}
where the cell boundaries \(m_{j\pm1/2}\) of the \(j\)-th cell are defined by
\begin{equation*}
	m_{j} = 0.5 \left(m_{j+1/2} + m_{j-1/2}\right).
\end{equation*}
\(\Sigma_{\text{d},j}\) corresponds to the dust surface density of the dust in the \(j\)-th mass bin, which clearly depends on the bin width. To be able to compare model results between models with different mass resolutions, it is more convenient to work with the quantity \(\sigma_\text{d}\) (defined in Eq. \eqref{eq:def_sigma_d}), which does not depend on the bin width.

In the discretized case, Eq. \eqref{eq:Sigma_ij} can be written as
\begin{equation}
	\Sigma_{\text{d},j} = \sigma_\mathrm{d}(r, m_j) \Delta,
\end{equation}
with the constant mass-weighted bin width 
\begin{align*}
\Delta = \frac{m_{j+1/2} - m_{j-1/2}}{m_j}.
\end{align*}
With \(A=m_{j+1/2}/{m_{j-1/2}} = m_{j+1} / m_j\), which is the same for all \(j\), because the grid is strictly logarithmic, one obtains
\begin{align}
\sigma_{\mathrm{d},j}(r)\equiv \sigma_\mathrm{d}(r,m_j) = \Sigma_{\text{d},j}(r)\frac{A+1}{2(A-1)}.
\end{align}

\subsection{Mean Mass}
\label{sec:appendix_mean_width}
The mean mass is given by
\begin{equation}
	\langle m\rangle= \frac{1}{\Sigma_\text{d}}\sum_j\Sigma_{\text{d},j} m_j,
\end{equation}
with the total dust surface density
\begin{equation}
	\Sigma_\text{d} = \sum_j \Sigma_{\text{d},j}.
\label{eq:total_sigmad}
\end{equation}
We set \(\Sigma_{\text{d},j}=0\) for \(\Sigma_{\text{d},j} < \Sigma_{\text{floor},j}\), where \(\Sigma_{\text{floor},j}(r_i) = m_j / A_{\text{cell},i}\), with the surface area \(A_{\text{cell},i}\) of the \(i\)-th radial grid cell. 

\subsection{Mean and maximum Stokes numbers}
The mean Stokes number is computed by
\begin{equation}
	\langle \mathrm{St} \rangle= \frac{1}{\Sigma_\text{d}}\sum_j\Sigma_{\text{d},j} \mathrm{St}_j,
\end{equation}
where \(\mathrm{St}_j\) is the Stokes number of the \(j\)-th mass bin. We define the maximum Stokes number as the Stokes number for which 99\% of the dust mass is constituted by particles with smaller Stokes numbers.

\section{Estimation of growth barriers}
\label{sec:appendix_barriers}
In the following, we derive estimates for the masses corresponding to the growth barriers using the expressions for the Stokes numbers presented in Section \ref{sec:growth_barriers}.

\subsection{Drift barrier}
In the case of the SC models, we can solve Eq. \eqref{eq:St_d} for the mass. This yields
\begin{align}
	\label{eq:m_d}
	\nonumber 
	m_\mathrm{d} = \min&\left[m_\bullet \left(\frac{2\Sigma_\mathrm{d}v_K^2\left|\frac{\mathrm{d}\ln P}{\mathrm{d}\ln r}\right|^{-1}}{\pi \rho_\mathrm{s}a_\bullet c_\mathrm{s}^2}\right)^{D/(D-2)}\right. ,\\
	 &\left. m_\bullet \left(\frac{2\Sigma_\mathrm{d}v_K^2\left|\frac{\mathrm{d}\ln P}{\mathrm{d}\ln r}\right|^{-1}}{\pi \rho_\mathrm{s}a_\bullet c_\mathrm{s}^2}\right)^3\phi_\text{min}^{-2}\right],
\end{align}
where we assumed that the drifting particles are in the Epstein regime.
The first expression inside the \(\min\) function corresponds to the case that the drift barrier is within the fractal growth regime, while the second expression corresponds to the case that the drift barrier is within the non-fractal growth regime.

\subsection{Fragmentation barrier}
As for the drift barrier, we can solve Eq. \eqref{eq:St_f} for the mass in the case of the SC models. However, now we have to consider both the Epstein and the Stokes regime, because the fragmentation barrier can lie in the Stokes regime in the inner parts of the disk. The fragmentation mass is given by
\begin{equation}
	\label{eq:m_f}
	m_\text{f} = \min\left(m_\mathrm{f}^\mathrm{Ep.}, m_\mathrm{f}^\mathrm{St.I}\right),
\end{equation}
where the fragmentation mass in the Epstein regime is given by
\begin{align}
	\nonumber 
	m^\text{Ep.}_\text{f} = \min\left[m_\bullet\left(\frac{2}{3\pi}\frac{v_\text{f}^2}{\alpha c_\mathrm{s}^2}\frac{\Sigma_\mathrm{g}}{\rho_\mathrm{s} a_\bullet}\right)^{D/(D-2)},\right.\\
	\left. m_\bullet\left(\frac{2}{3\pi}\frac{v_\text{f}^2}{\alpha c_\mathrm{s}^2}\frac{\Sigma_\mathrm{g}}{\rho_\mathrm{s} a_\bullet}\right)^3\phi_\text{min}^{-2}\right],
	\label{eq:m_f_Ep}
\end{align}
and the fragmentation mass in the Stokes regime is given by
\begin{align}
	\nonumber
	m^\text{St.I}_\text{f} = \min\left[m_\bullet \left(\frac{3}{2\pi}\frac{v_\text{f}^2 \lambda_\text{mfp}\Sigma_\mathrm{g}}{a_\bullet^2\rho_\mathrm{s}\alpha c_\mathrm{s}^2}\right)^{D/(D-1)}\right.,\\
	\left. m_\bullet \left(\frac{3}{2\pi}\frac{v_\text{f}^2 \lambda_\text{mfp}\Sigma_\mathrm{g}}{a_\bullet^2\rho_\mathrm{s}\alpha c_\mathrm{s}^2}\right)^{3/2}\phi_\text{min}^{-1/2}\right].
\end{align}

\subsection{Bouncing Barrier}
Typically, the bouncing barrier is well within the Epstein regime. Since the bouncing-limited mass distribution is rather narrow, we assume equal-sized collisions \citep{DominikDullemond2024}. For the SC models, we can solve the first expression in the brackets of Eq. \eqref{eq:St_b} for the mass, which yields
\begin{align}
	\nonumber
	m_\text{b0} = \min\left[m_\bullet^{(D-2)/(2D-2)}\left(\frac{20}{3}\frac{\Sigma_\mathrm{g} F_\text{roll}}{\alpha c_s^2\rho_s}\right)^{D/(2D-2)}\right. ,\\
	\left. m_\bullet^{1/4}\phi_\text{min}^{-1/2}\left(\frac{20}{3}\frac{\Sigma_\mathrm{g} F_\text{roll}}{\alpha c_s^2\rho_s}\right)^{3/4}\right].
	\label{eq:m_b0}
\end{align}
Solving the second expression \(\mathrm{Re}^{-1/2}\) in the brackets of Eq. \eqref{eq:St_b} for the mass yields
\begin{align}
	\nonumber
	m_\eta = \min\left[m_\bullet\left(\frac{2\Sigma_\text{g}}{\pi\rho_s a_\bullet\sqrt{\mathrm{Re}}}\right)^{D/(D-2)}\right.,\\
	\left. m_\bullet\phi_\text{min}^{-2}\left(\frac{2\Sigma_\text{g}}{\pi\rho_s a_\bullet\sqrt{\mathrm{Re}}}\right)^3\right].
	\label{eq:m_eta}
\end{align}

Thus, the final expression for the mass corresponding to the bouncing barrier is given by
\begin{equation}
	\label{eq:m_b}
	m_\text{b} = \max\left(m_\eta, m_\text{b0}\right),
\end{equation}

The expressions of Eqs. \eqref{eq:m_f_Ep} and \eqref{eq:m_b0} are similar those of \citet{Estrada2022}, but there is a small difference due to a slightly different definition of the Stokes number.

\subsection{Lower bouncing barrier}
The smallest particles of the size distribution can only grow, if the condition
\begin{equation}
	v_\text{rel}(m_\text{lb},m) \geq v_\text{b}(m_\text{lb},m) \quad\text{for } m_\text{lb} \leq m
\end{equation}
is met, where \(m_\text{lb}\) is the lower edge of the mass distribution.
Neglecting the case, where \(v_\mathrm{f} < v_\mathrm{s}\), the bouncing velocity is approximately given by
\begin{equation}
	\label{eq:approx_vbounce}
	v_\text{b} \approx \sqrt{\frac{k}{m_\text{lb}}},
\end{equation}
with \(k=5\pi a_\bullet F_\mathrm{roll}\).
For Eq. \eqref{eq:approx_vbounce}, we exploited the fact that we are only interested in the range \(m\geq m_\mathrm{lb}\), such that we can approximate \(\left(m_\mathrm{lb} + m\right)/\left(m_\mathrm{lb}m\right)\approx1/m_\mathrm{lb}\). The deviation from the exact expression is at maximum a factor of 2.

To find reasonable approximations for the relative velocities, we need to distinguish between the case, where \(D\) is 2 or close to 2, and the case, where \(D\) is considerably different from 2. 

For the case, where \(D\) is significantly different from 2, we assume that relative velocities are dominated by turbulence and that the particles are well within the \(t_\mathrm{s} < t_\eta\) regime (i.e., \(\mathrm{St}\ll\mathrm{Re}^{-1/2}\)). Furthermore, we assume that the Stokes number \(\mathrm{St}\) of the collision partner is considerably larger than the Stokes number \(\mathrm{St_{lb}}\) of the particle at the lower edge of the mass distribution. With these assumptions, the relative velocity can be approximated by 
\begin{equation}
	v_\mathrm{rel}^2 \approx \frac{3}{2}\alpha c_\mathrm{s}^2\mathrm{St}^2\sqrt{\mathrm{Re}}.
\end{equation}
Since \(D\) is significantly different from 2, the Stokes number and therefore the relative velocity increases with mass. Hence, we know that
\begin{equation}
	\label{eq:v_rel_geq}
	v_\mathrm{rel}^2 \geq \frac{3}{2}\alpha c_\mathrm{s}^2\mathrm{St_{lb}}^2\sqrt{\mathrm{Re}}.
\end{equation}
The right-hand side of Eq. \eqref{eq:v_rel_geq} is the lower bound of the relative velocities, which we can compare to the approximation of the bouncing velocity given by Eq. \eqref{eq:approx_vbounce}. This yields 
\begin{equation}
	\mathrm{St_{lb}} = \frac{v_\mathrm{b}\mathrm{Re}^{-1/4}}{\sqrt{\frac{3}{2}\alpha}c_\mathrm{s}} \quad \text{for } D\not\approx2.
\end{equation}
We can again solve for the mass, which yields
\begin{align}
	\label{eq:m_lb}
	m_\text{lb} = \left(\frac{2k\text{Re}^{-1/2}}{3\alpha c_s^2\chi^2}\right)^{D/(3D-4)} \quad \text{for } D\not\approx2,
\end{align}
with \(\chi = \pi\rho_s a_\bullet m_\bullet^{2/D-1} /(2\Sigma_\mathrm{g})\).

Now, we consider the case, where \(D\approx2\). We define \(v_0\) as the contribution of all sources of relative velocities except Brownian motion:
\begin{equation}
	v_0 = \sqrt{\left(v_\text{rel}^\text{turb}\right)^2 + \left(v_\text{rel}^\text{rad}\right)^2
	+ \left(v_\text{rel}^\text{azi}\right)^2 + \left(v_\text{rel}^\text{vert}\right)^2}.
	\label{eq:v0}
\end{equation}
For \(D\!=\!2\), \(v_0(m_1, m_2) \!=\! v_0(m_\bullet, m_\bullet)\). In other cases, where \(v_0\) slightly depends on mass, we neglect this mass dependence and approximate \(v_0(m_1, m_2) \!\approx\! v_0(m_\bullet, m_\bullet)\).

Since the Stokes number only weakly depends on mass for \(D\approx2\), \(v_0\) is approximately constant. The contribution of Brownian motion quickly decreases with mass, therefore, we can assume that \(v_0\) dominates in the parts of the mass domain that we are interested in. Comparing \(v_0\) to the approximation of the bouncing velocity given by Eq. \eqref{eq:approx_vbounce} yields
\begin{equation}
	m_\mathrm{lb} = \frac{k}{v_0^2} \quad \text{for } D\approx2.
\end{equation}

\subsection{AC models and low turbulence case}
Since the estimates of the growth barriers are plotted as the contours where \(\mathrm{St}= \mathrm{St_{d/f/b/lb}}\) instead of using the mass expressions (except the lower bouncing barrier for \(D\!\lesssim2\)), they do not have to be adjusted for the AC models. For \(D \!\lesssim\! 2\), there is no Stokes expression for the lower bouncing barrier. Thus, we apply the mass expression Eq. \eqref{eq:m_lb}. Since this was derived for the \(D=2\) SC models, it shows some deviations from the numerical results of the AC models.

For the low turbulence case however, the assumption that turbulence is the dominating source of relative velocity does not hold anymore. For the fragmentation and lower bouncing barrier (for \(D \nlesssim 2\)), we therefore apply adjusted estimates by approximating \(v_\mathrm{rel}^\mathrm{rad}\approx 2v_\mathrm{drift}^\mathrm{max}\mathrm{St}\). This yields 
\begin{equation}
	\mathrm{St}_\text{f,lt} = \frac{v_\text{f}}{2v_\text{drift}^\text{max}} \quad \text{and } \mathrm{St_{lb,lt}} = \frac{v_\mathrm{b}}{2v_\text{drift}^\text{max}}, 
\end{equation}
respectively.
\section{Dust distribution for high turbulence}
\label{sec:appendix_plots}
Here, we present the dust distribution plot after \(10^6\) years of the high turbulence (\(\alpha=10^{-3}\)) models with bouncing. It becomes clear that the fragmentation and bouncing barrier are very close together for \(D=3\), such that some collision result in fragmentation although the bouncing barrier is active. For \(D=2.5\) the same is true in the outer parts of the disk, while in the inner parts of the disk fragmentation events do not occur. For \(D=2\), fragmentation events do not play a role in any part of the disk.

\begin{figure}
   \centering
   \includegraphics[width=0.8\hsize]{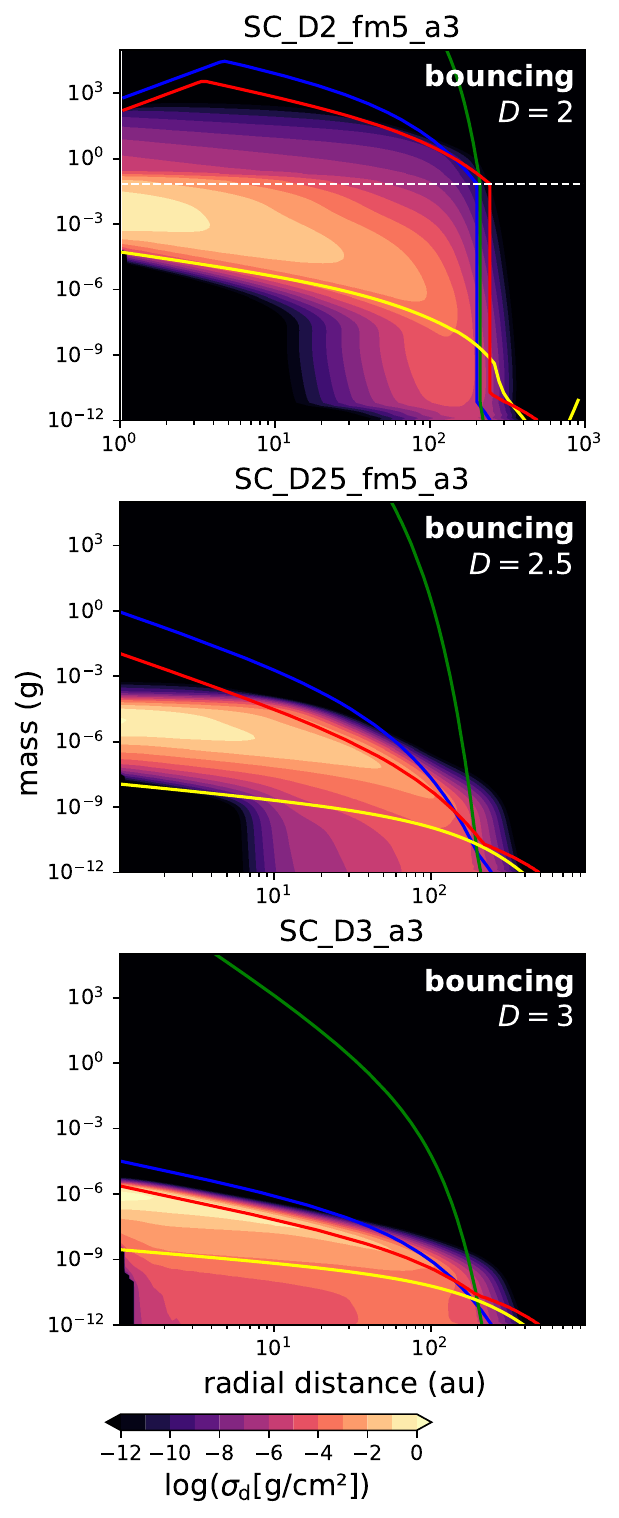}
      \caption{Same as Fig. \ref{fig:sc_sigma_d_varyD} but for high turbulence models (\(\alpha=10^{-3}\)) with bouncing.}
         \label{fig:sc_sigma_d_highturb}
\end{figure}

\end{appendix}
\end{document}